# Advancing Network Securing Strategies with Network Algorithms for Integrated Air Defense System (IADS) Missile Batteries

**Rakib Hassan Pran**
**M.Sc. in Applied Statistics with Network Analysis**
**International Laboratory for Applied Network Research**
**Research Departments of NRU HSE**
**National Research University Higher School of Economics**
**RU/Российская Федера́ция**

*Abstract*— **Recently, the Integrated Air Defense System (IADS) has become vital for the defense system as the military defense system is vital for national security. Placing Integrated Air Defense System batteries among locations to protect locations assets is a crucial problem because optimal solutions are needed for interceptor missiles to intercept attacker missiles for maximizing protection of assets across locations or places. In this research, the procedures of using network algorithms along with developing several network algorithms are going to be demonstrated to develop a model for sequential development of seven network securing strategies of placing Surface to Air Missile (SAM) batteries to maximize the protection of assets across locations (based on given asset values) by generating optimal solutions through computation to destroy maximum attacker missiles by using minimum interceptor missiles with given intercept probability. This network securing strategies can be implemented not only for Integrated Air Defense System (IADS) planning but also Counter Air (CA) planning as Integrated Air Defense System (IADS) is conducted with defensive counter air supported by attack operations in offensive counter air.**

## I. INTRODUCTION

Integrated Air Defense System (IADS) is an air defense system which is the aggregation of Service/functional component and agency Air Missile Defense (AMD) systems consist of sensors, weapons, C2, communications, intelligence systems and personnel operating in a theater or Joint Operations Area (JOA) [1]. Generally, IADS is established by Area Air Defense Commander (AADC) and, a theater AMD system in IADS typically depends on support and enabling functions from national assets and systems which are not controlled by the JFC (Joint Force Commander) [1]. Air Defense Systems are designed to protect task forces, important military and state installations, to counter mass raids of air attack weapons, attack of aviations including modern and prospective aircrafts, air systems, strategic cruise missiles, aeroballistics, tactical and theatre ballistic missiles, remotely piloted vehicles, etc, within varieties of operational altitudes with static or dynamic positions under conditions of mass attack, Electronic Counter Measures (ECM) conditions, fire counteraction environment conditions, etc [2]. As part of IADS functionality, IADS needs to provide protection of a country's assets by placing IADS missile batteries in optimal locations across the country [3]. An IADS missile battery is usually an unit of AMD which can be manually portable such as tactical air defense weapons systems or unportable such as emplaced IADS missile batteries to protect population over places which are not mobile and readily detectable based on the functionality, structure and attribute of IADS missile battery [3]. Placing an IADS missile battery over locations is a critical problem as limited IADS missile batteries are intended to protect assets over a large number of locations by considering missile ranges of IADS missile batteries for asset values over locations [3]. Locations can be represented in a network as locations are connected through real world routes or paths and on the other hand, locations can also be represented as a network where edges represent euclidean distances among locations. But in case of missile range, euclidean distances among locations need to be considered as edges for representing a location network [3]. In case of placing IADS missile batteries in a network, optimal location nodes need to be found, for that, a huge number of network algorithms existed in literature such as network algorithms based on the sum of the fraction of all-pairs shortest paths , the reciprocal of the average shortest path distance, the structure of incoming links, the centrality of neighbors, the sum of the fraction of all-pairs shortest paths, etc [4-10].

In this research, a model for sequential development of seven strategies with the use of six network algorithms has been introduced. These seven strategies have been applied to generated weighted edge location networks with assigned asset values. These weighted edge location networks with generated assigned values are basically generated by using Watts Strogatz graph algorithm [11] which is based on small worldness [12] computed with network clustering coefficient algorithm [13] [14]. The results that came from after applying seven strategies to generated location networks have been analyzed with ordinary least square linear regression analysis [15] at the end.

## II. Methodology and Computational Experiment

For developing sequential model of seven network securing strategies, six network algorithms have been selected to measure each node of the location networks, where network_algorithm_1 is based on the sum of the fraction of all-pairs shortest paths, network_algorithm_2 is based on the reciprocal of the average shortest path distance, network_algorithm_3 is a ranking of nodes in graph based on the structure of incoming links, network_algorithm_4 is based on the centrality of neighbors, network_algorithm_5 is based on the sum of the fraction of all-pairs shortest paths similar to network_algorithm_1, and network_algorithm_6 is measuring a node as the fraction of nodes that it is connected to [4-10]. These network algorithms have been used to develop a sequential developing model of seven network securing strategies in the section of IV named as "STRATEGIES".

To generate weighted edge location networks with assigned asset values, Watts Strogatz small world network generation [11] [12] algorithm has been selected where assigned asset values (see Figure 4) are following power law distribution [3]. Besides considering Euclidean distances, it is assumed that to hit each location's attacker missile, the interceptor missiles have different flying altitudes and different trajectory paths which doesn't depend on Euclidean distance because of geographical attributes such as hills, mountains, etc. For that reason, the trajectory path distance of an interceptor missile has been considered as range for that interceptor missile. Because of considering different flying altitudes' trajectory paths as distances among nodes, generated location networks don't follow Pythagorean theorem or Euclidean theorem [28][29].

As a scale of small worldness [12], five types of small worldness have been selected for generating 20 networks of 50 location nodes with different diameters where small worldness is close to zero means more to small world characteristics.

By applying seven strategies upon previously generated 20 location networks with 50 nodes, total unprotected asset value percentages (Worst Case Scenarios) have been computed for each strategy with 98% of IADS missile battey's interceptor missile's interceptor probability [16]. To analyze the result that came from applying seven strategies upon generated location networks, ordinary least square linear regression analysis [15] has been used for analyzing relations among sum of all IADS missile ranges, network diameters of generated location networks and unprotected asset value percentages for optimal strategy of each generated network. Besides that, ordinary least square linear regression analysis [15] has also been used for analyzing relations among sum of all IADS missile ranges, network diameters of generated location networks and small worldness for optimal strategy of each generated network.

## III. Computational Instruments

Python 3 Google Compute Engine [17] has been used where System RAM is 12.7 GB and Disk is 107.7 GB. Code has been written in Python language [18] and Packages are: Google.colab: 0.0.1a2, Networkx: 3.0 [19], Pandas: 1.3.5 [20], Numpy: 1.21.6 [21], Scipy: 1.10.0 [22], Matplotlib: 3.5.3 [23], Array, Statsmodels.api: 0.13.5 [24], Sklearn: 1.2.1 [25]. Documentation done in Google Docs [30].

## IV. Strategies

In this section, seven network securing strategies have been developed with six network algorithms which have been introduced in the Methodology section.

**Strategy_1:**

Step_1. Considering weighted edge network with asset values for placing IADS's batteries

Step_2. Computing network_algorithm_1

Step_3. Putting IADS missile battery with maximum interceptor missile range on the place of first maximum network_algorithm_1 measure and if second IADS missile battery exists with with maximum interceptor missile range or less than interceptor missile range then, putting the second IADS missile battery on the place of the next maximum network_algorithm_1 measure where distance from the node of first maximum network_algorithm_1 measure is more than maximum IADS missile battery interceptor missile range.

Step_4. And doing the same for placing rest of IADS batteries on next maximum network_algorithm_1 measure where node distance from first placed IADS missile battery is more than first IADS battery's interceptor missile range

For example, one generated weighted edge location network has been introduced below. One, two and three IADS missile batteries have been applied sequentially on that generated weighted edge location network where interceptor missile ranges are sequentially 80 km, 70 km and 20 km.

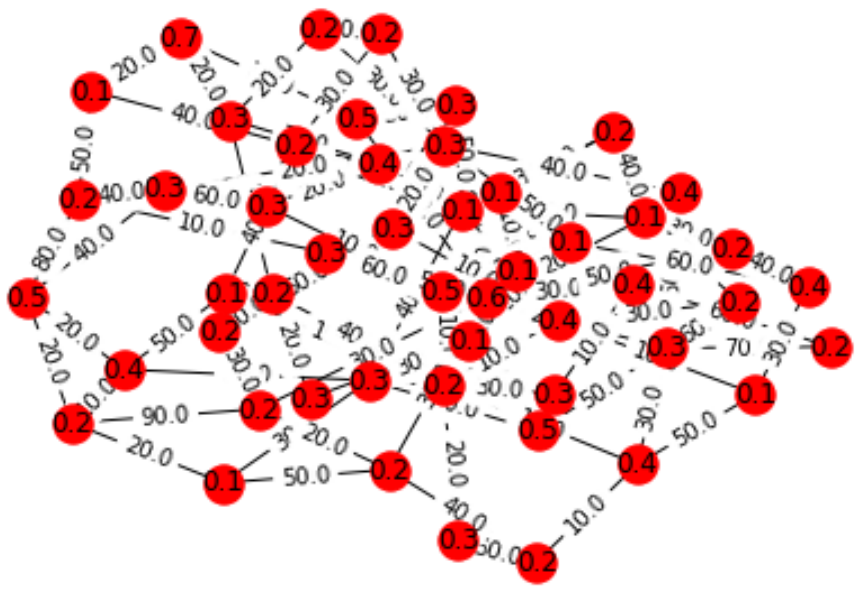

**Figure 1: Weighted edge network with asset values. Each asset value is scored as a fraction number (between 0.0 and 1.0) out of 1. All nodes are unprotected (colored as red) as no IADS missile battery has been placed. Here, The diameter of network is 180 km**

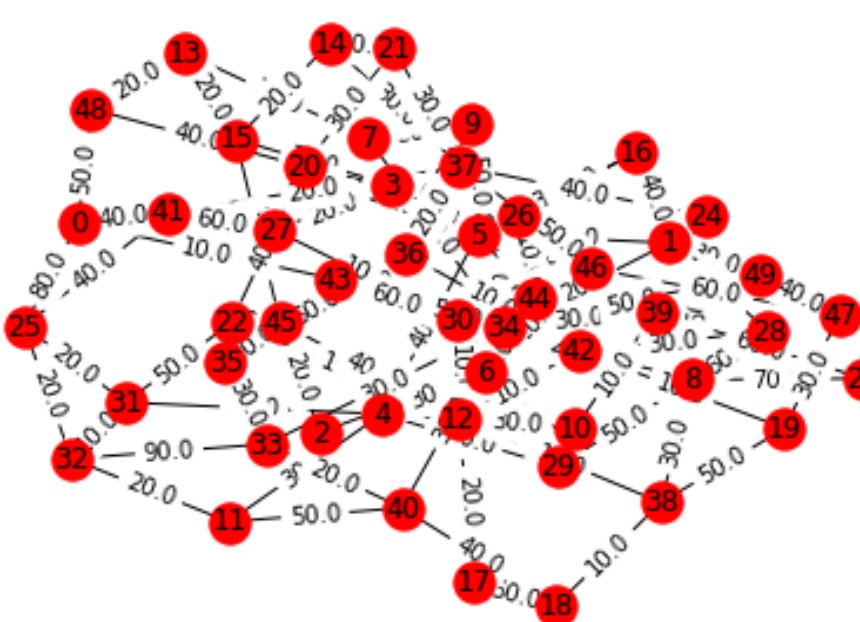

**Figure 2: Weighted edge network with location node numbers. All nodes are unprotected (colored as red) as no IADS missile battery has been placed. Here, The diameter of network is 180 km**

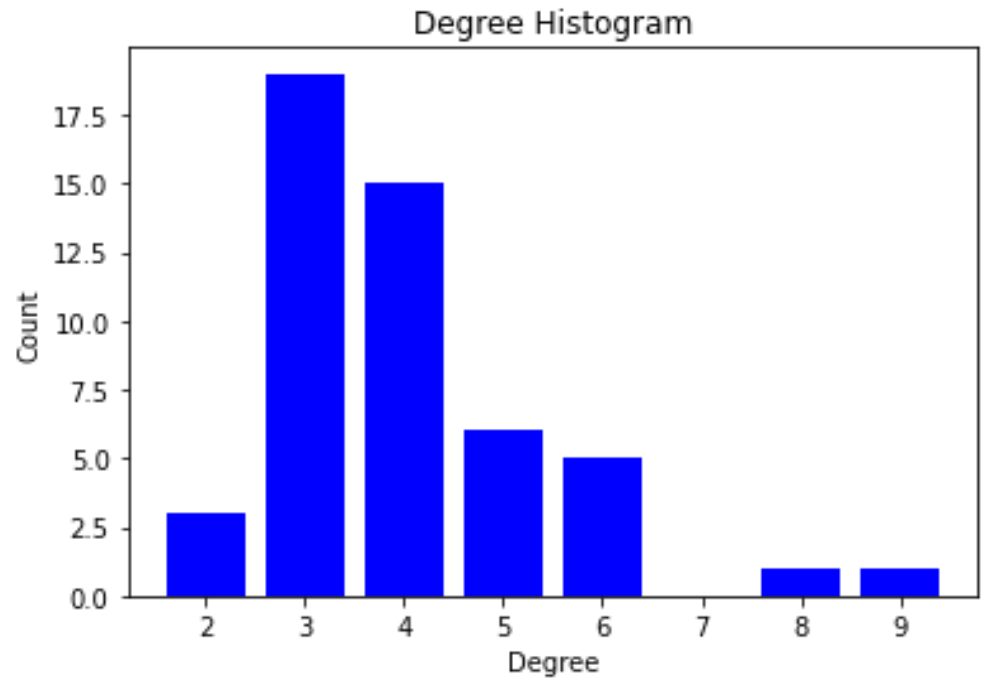

**Figure 3: Degree distribution histogram of weighted edge network**

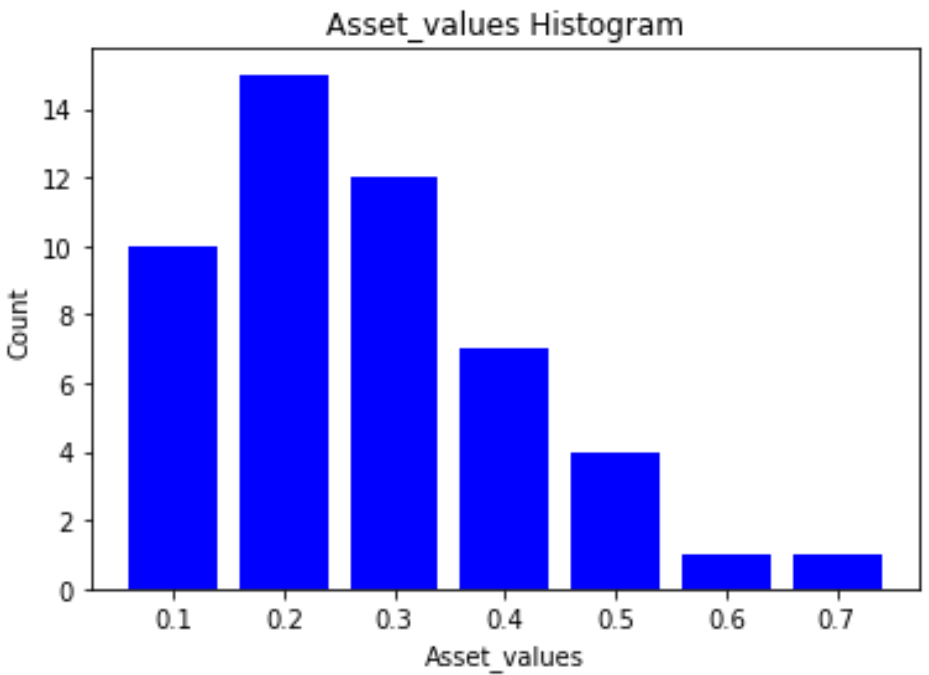

**Figure 4: Asset value distribution histogram of weighted edge network**

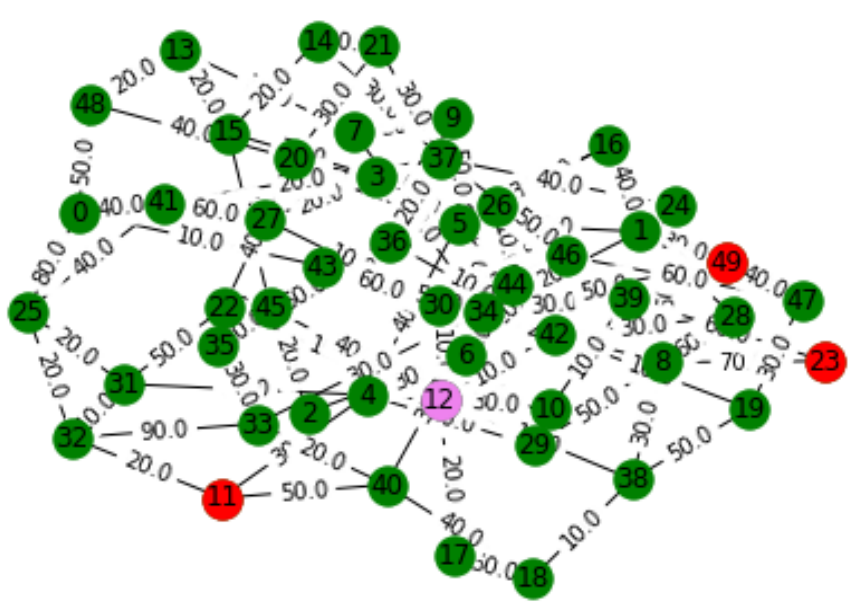

**Figure 5: Violet nodes are where IADS missile batteries have been placed and red nodes are unprotected location node numbers where IADS missile battery's interceptor missiles can't reach. One IADS missile battery is placed at location node number 12. Placed IADS missile battery's interceptor missile range is 80 km Here, total unprotected asset value is 0.5**

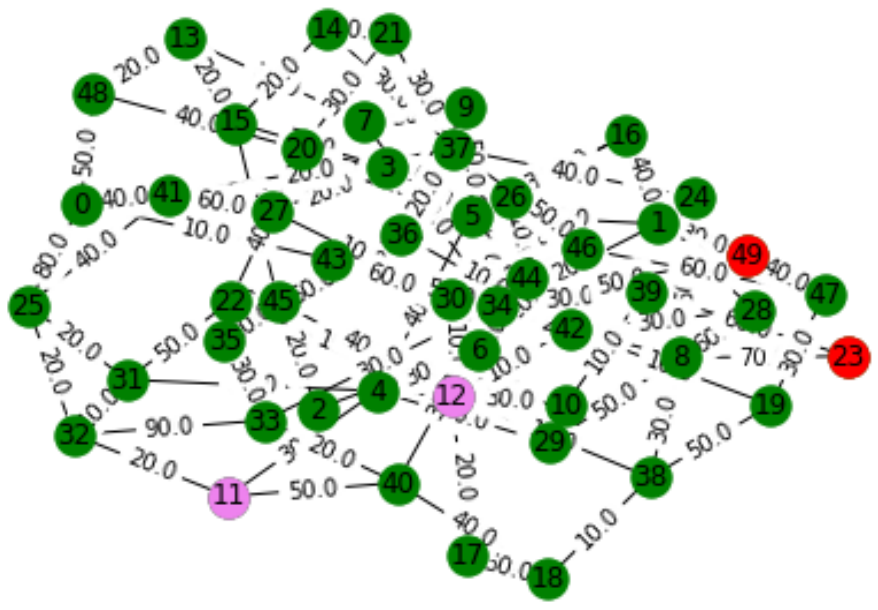

**Figure 6: Violet nodes are where IADS missile batteries have been placed and red nodes are unprotected location node numbers where IADS missile battery's interceptor missiles can't reach. First IADS missile battery is placed at location node number 12 and Second IADS missile battery is placed at location node number 11. First placed IADS missile battery's interceptor missile range is 80 km and second placed IADS missile battery's interceptor missile range is 70 km. Here, total unprotected asset value is 0.4**

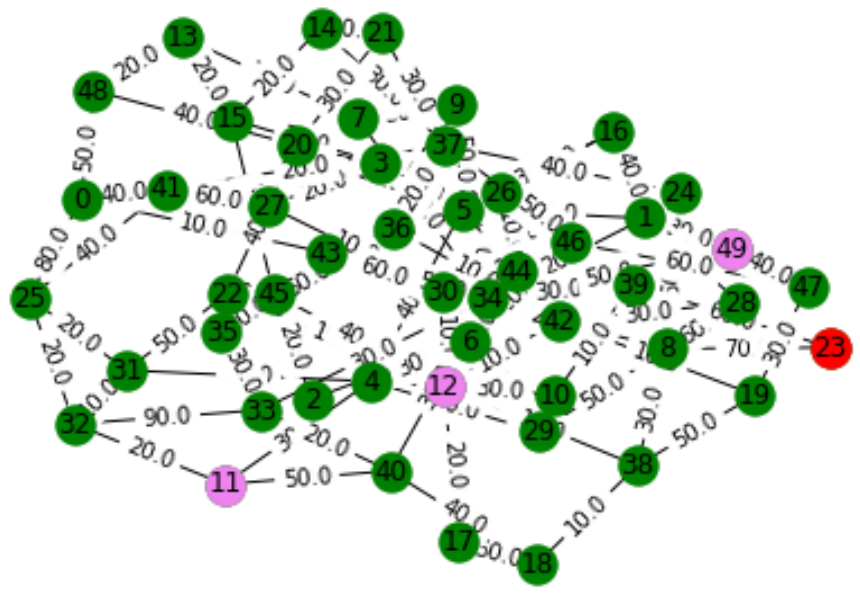

**Figure 7: Violet nodes are where IADS missile batteries have been placed and red nodes are unprotected location node numbers where IADS missile battery's interceptor missiles can't reach. First IADS missile battery is placed at location node 12, Second IADS missile battery is placed at location node 11 and Third IADS missile battery is placed at location node 49. First placed IADS missile battery's interceptor missile range is 80 km, second placed IADS missile battery's interceptor missile range is 70 km and third placed IADS missile battery's interceptor missile range is 20 km. Here, total unprotected asset value is 0.2**

### Strategy_2:

(same as Strategy_1 except using network_algorithm_2 instead of network_algorithm_1)

Step_1. Considering weighted edge network with asset values for placing IADS's batteries

Step_2. Computing network_algorithm_2

Step_3. Putting IADS missile battery with maximum interceptor missile range on the place of first maximum network_algorithm_2 measure and if second IADS missile battery exists with with maximum interceptor missile range or less than interceptor missile range then, putting the second IADS missile battery on the place of the next maximum network_algorithm_2 measure where distance from the node of first maximum network_algorithm_2 measure is more than maximum IADS missile battery interceptor missile range.

Step_4. And doing the same for placing rest of IADS batteries on next maximum network_algorithm_2 measure where node distance from first placed IADS missile battery is more than first IADS battery's interceptor missile range

For example, One, two and three IADS missile batteries have been applied sequentially on that previously generated weighted edge location network where interceptor missile ranges are sequentially 80 km, 70 km and 20 km.

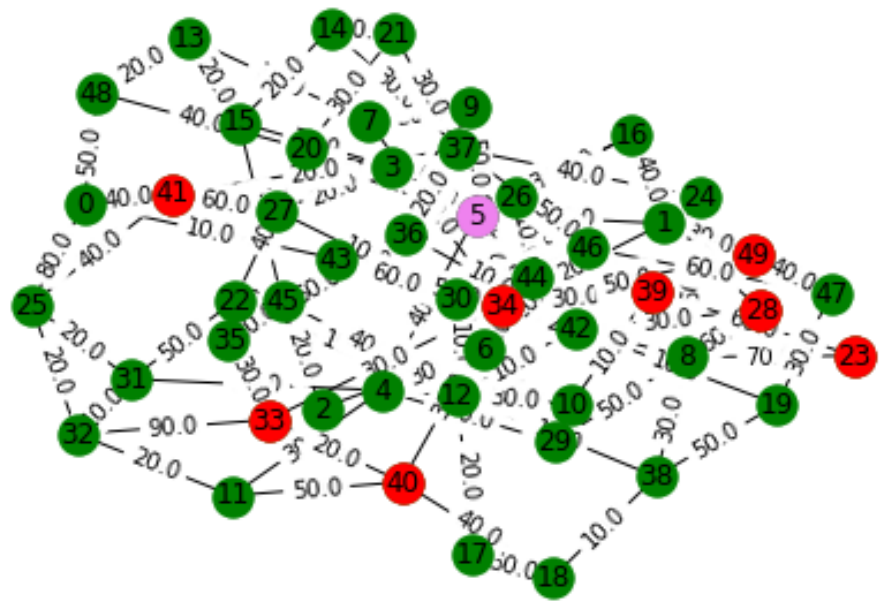

**Figure 8: Violet nodes are where IADS missile batteries have been placed and red nodes are unprotected location node numbers where IADS missile battery's interceptor missiles can't reach. One IADS missile battery is placed at location node number 5. Placed IADS missile battery's interceptor missile range is 80 km Here, total unprotected asset value is 2.3000000000000003**

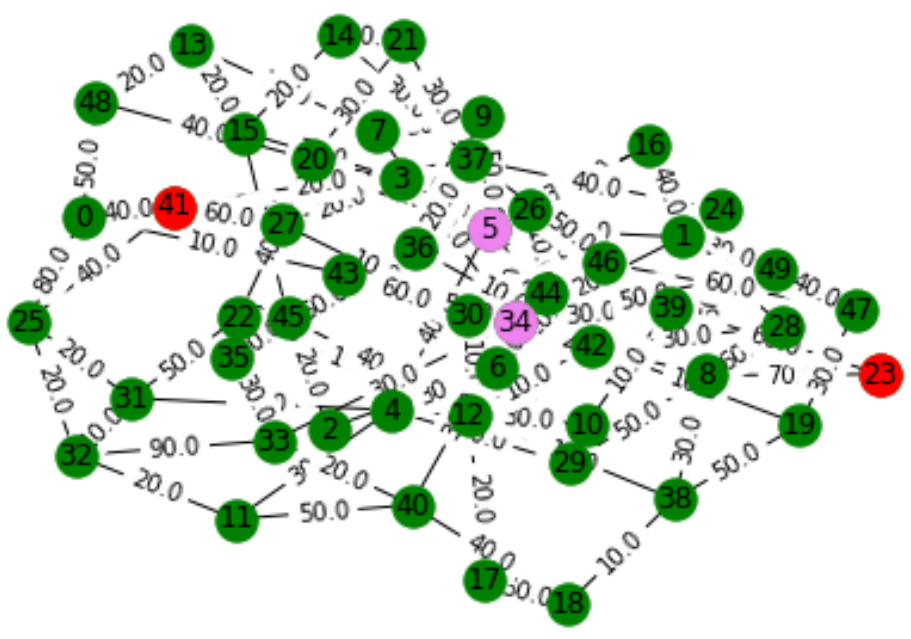

**Figure 9: Violet nodes are where IADS missile batteries have been placed and red nodes are unprotected location node numbers where IADS missile battery's interceptor missiles can't reach. First IADS missile battery is placed at location node 5 and Second IADS missile battery is placed at location node 34. First placed IADS missile battery's interceptor missile range is 80 km and second placed IADS missile battery's interceptor missile range is 70 km. Here, total unprotected asset value is 0.5**

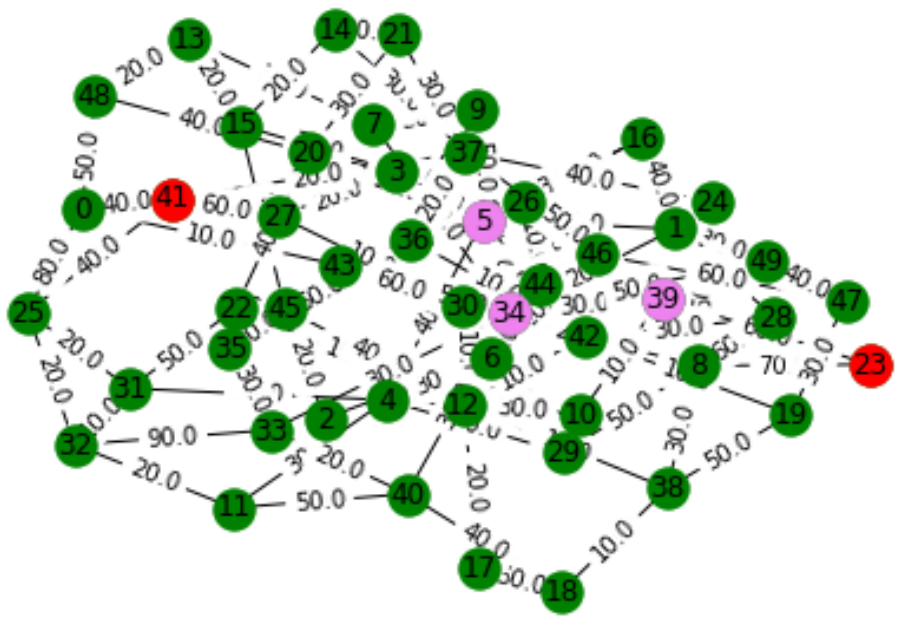

**Figure 10: Violet nodes are where IADS missile batteries have been placed and red nodes are unprotected location node numbers where IADS missile battery's interceptor missiles can't reach. First IADS missile battery is placed at location**

**node 5, Second IADS missile battery is placed at location node 34 and Third IADS missile battery is placed at location node 39. First placed IADS missile battery's interceptor missile range is 80 km, second placed IADS missile battery's interceptor missile range is 70 km and third placed IADS missile battery's interceptor missile range is 20 km. Here, total unprotected asset value is 0.5**

### Strategy_3:

Step_1. Considering weighted edge network with asset values for placing IADS's batteries.

Step_2. Computing network_algorithm_1, network_algorithm_2, network_algorithm_3, network_algorithm_4, network_algorithm_5 and network_algorithm_6.

Step_3. Creating six sequences by ordering location node numbers as descending orders of the results of six network algorithms sequentially such as sequence_number_1 is the sequence of node numbers ordered as descending order of network_algorithm_1's measuring values on nodes.

Step_4. Estimating total unprotected asset values for placing IADS missile battery with maximum missile range on the place of first location node number from each sequence and choosing the lowest unprotected asset value corresponded node among six unprotected asset value corresponded nodes for placing first IADS with maximum missile range.

Step_5. Estimating total unprotected asset values for placing second IADS missile battery (if any) with maximum missile range or with second maximum missile range on the place of second location node number from each sequence and choosing the lowest unprotected asset value corresponded location node among six unprotected asset value corresponded location nodes for placing second IADS with maximum missile range or with second maximum missile range.

If the lowest unprotected asset value corresponded location node resulted from considering second node position of each sequence is same as the lowest unprotected asset value corresponded node resulted from considering first node position of each sequence, then repeat step 4 for the place of third location node number in sequences for placing second IADS missile battery with maximum missile range or with second maximum missile range.

Step_6. Repeating Step_4 and Step_5 for placing the next IADS missile battery with previous missile ranges or with the next longer missile range (which is less than previous missile range).

Step_7. If there are $M$ number of nodes where the lowest unprotected asset value corresponded location node resulted from considering any node position of each sequence is same as the lowest unprotected asset value corresponded node resulted from considering previous node position of each sequence, then, for placing the last $M$ number of IADS batteries with previous missile ranges or with the the most minimum missile ranges, detect all unprotected location nodes where previously placed IADS batteries interceptor missiles can't reach.

Step_8. Estimating total unprotected asset values for placing the last $M$ number of IADS batteries on each unprotected location node and choosing the lowest total unprotected asset value corresponding to $M$ number of unprotected location nodes for placing the last $M$ number of IADS batteries with previous missile ranges or with the most minimum missile ranges.

For example, One, two and three IADS missile batteries have been applied sequentially on that previously generated weighted edge location network where interceptor missile ranges are sequentially 80 km, 70 km and 20 km.

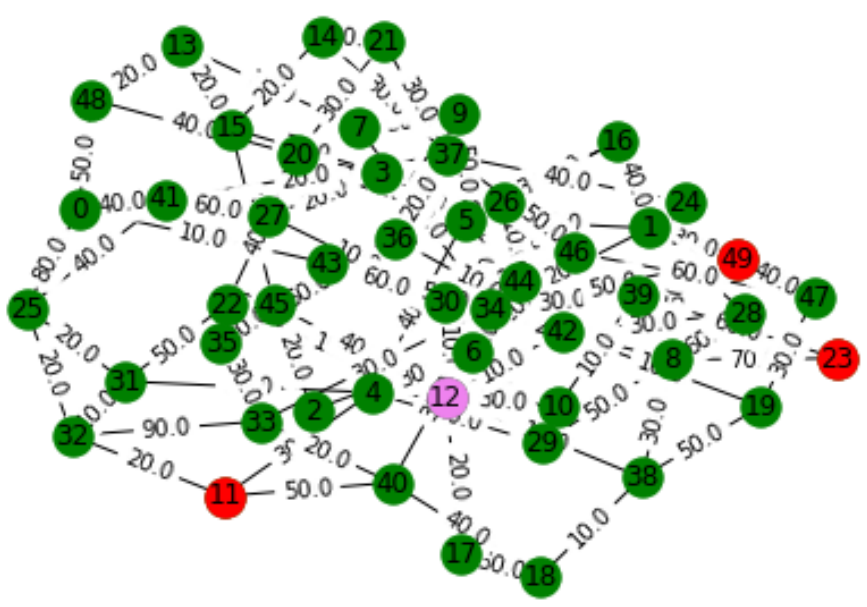

**Figure 11: Violet nodes are where IADS missile batteries have been placed and red nodes are unprotected location node numbers where IADS missile battery's interceptor missiles can't reach. One IADS missile battery is placed at location node number 12. Placed IADS missile battery's interceptor missile range is 80 km Here, total unprotected asset value is 0.5**

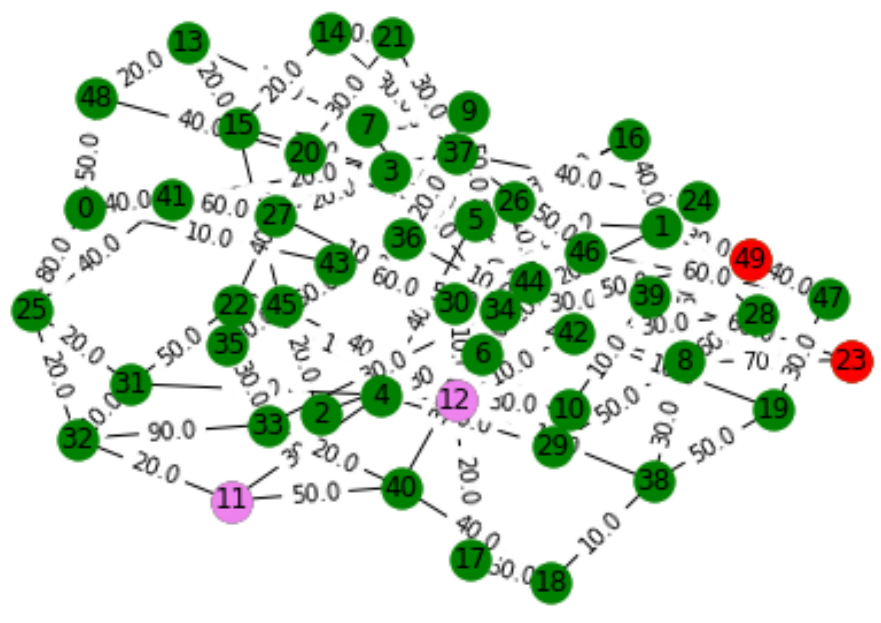

**Figure 12: Violet nodes are where IADS missile batteries have been placed and red nodes are unprotected location node numbers where IADS missile battery's interceptor missiles can't reach. First IADS missile battery is placed at location node number 12 and Second IADS missile battery is placed at location node number 11. First placed IADS missile battery's interceptor missile range is 80 km and second placed IADS**

**missile battery's interceptor missile range is 70 km. Here, total unprotected asset value is 0.4**

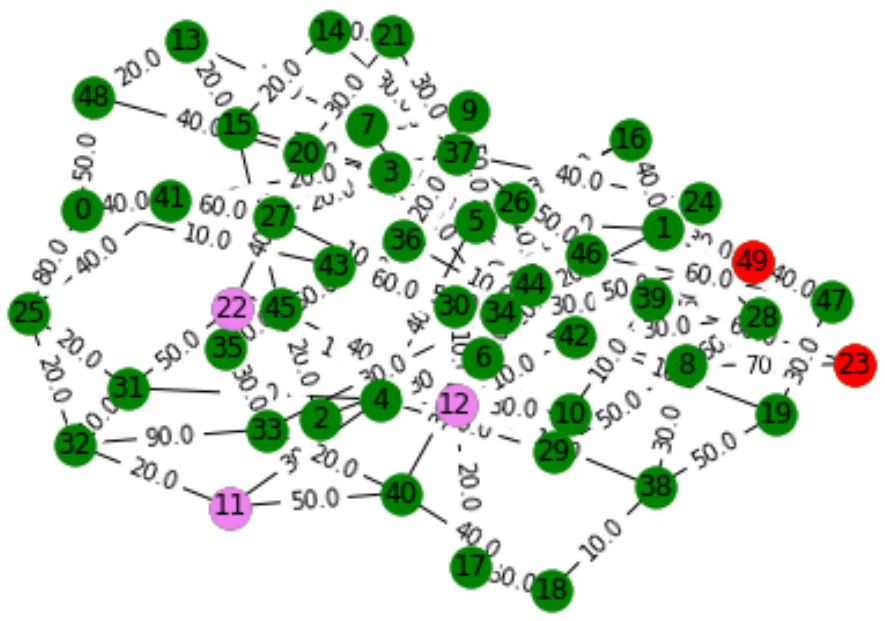

**Figure 13: Violet nodes are where IADS missile batteries have been placed and red nodes are unprotected location node numbers where IADS missile battery's interceptor missiles can't reach. First IADS missile battery is placed at location node 12, Second IADS missile battery is placed at location node 22 and Third IADS missile battery is placed at location node 11. First placed IADS missile battery's interceptor missile range is 80 km, second placed IADS missile battery's interceptor missile range is 70 km and third placed IADS missile battery's interceptor missile range is 20 km. Here, total unprotected asset value is 0.4**

## Strategy_4:

(same as Strategy_3 for first four steps)

Step_1. Considering weighted edge network with asset values for placing IADS's batteries

Step_2. Computing network_algorithm_1, network_algorithm_2, network_algorithm_3, network_algorithm_4, network_algorithm_5 and network_algorithm_6

Step_3. Creating six sequences by ordering location node numbers as descending orders of the results of six network algorithms sequentially such as sequence_number_1 is the sequence of node numbers ordered as descending order of network_algorithm_1's measuring values on nodes.

Step_4. Estimating total unprotected asset values for placing IADS missile battery with maximum missile range on the place of first location node number from each sequence and choosing the lowest unprotected asset value corresponded node among six unprotected asset value corresponded nodes for placing first IADS with maximum missile range.

Step_5. After placing the first IADS missile battery with maximum missile range, for placing the next IADS missile batteries, we shall consider those location nodes in six sequences where the distances from the first placed IADS missile battery are more than the first placed IADS missile battery's missile range.

Step_6. Repeating Step_4 for placing the next IADS missile battery with previous missile ranges or with the next longer missile range (which is less than previous missile range) until all IADS missile batteries are placed. If the lowest unprotected asset value corresponded location node resulted from considering any node position of each sequence is same as the lowest unprotected asset value corresponded node resulted from considering previous node position of each sequence, then, we shall consider the second lowest unprotected asset value corresponded location node resulted from considering that node position of each sequence for placing IADS missile battery.

For example, One, two and three IADS missile batteries have been applied sequentially on that previously generated weighted edge location network where interceptor missile ranges are sequentially 80 km, 70 km and 20 km.

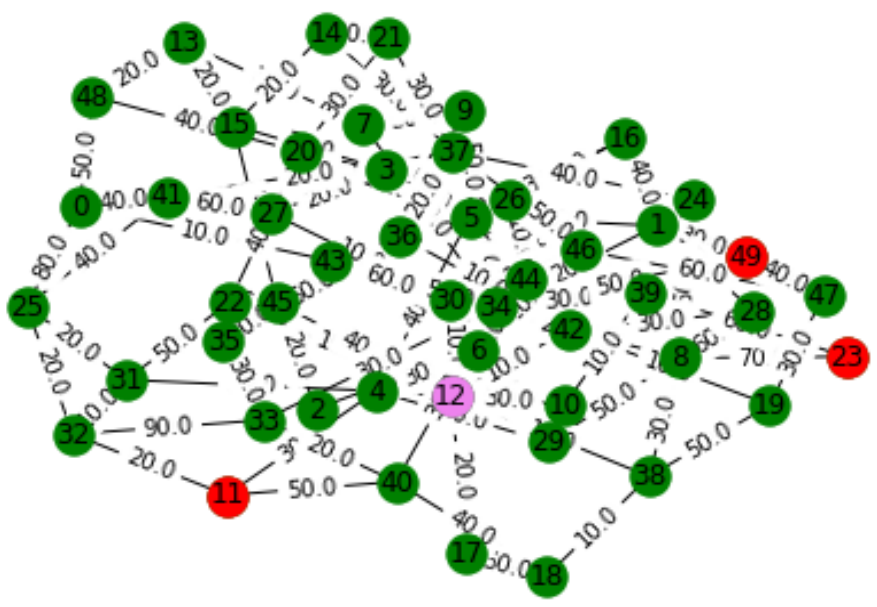

**Figure 14: Violet nodes are where IADS missile batteries have been placed and red nodes are unprotected location node numbers where IADS missile battery's interceptor missiles can't reach. One IADS missile battery is placed at location node number 12. Placed IADS missile battery's interceptor missile range is 80 km Here, total unprotected asset value is 0.5**

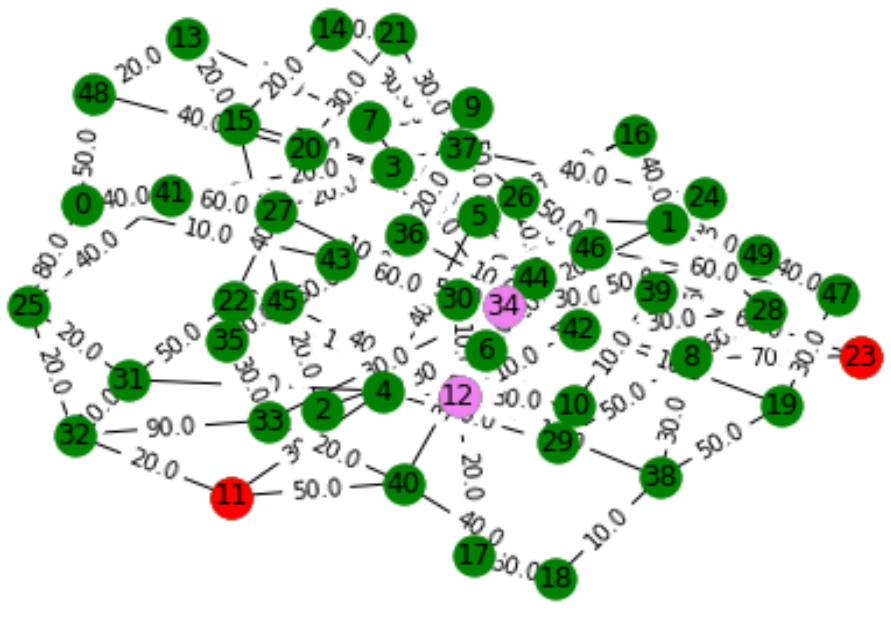

**Figure 15: Violet nodes are where IADS missile batteries have been placed and red nodes are unprotected location node numbers where IADS missile battery's interceptor missiles can't reach. First IADS missile battery is placed at location node number 12 and Second IADS missile battery is placed at location node number 34. First placed IADS missile battery's interceptor missile range is 80 km and second placed IADS**

**missile battery's interceptor missile range is 70 km. Here, total unprotected asset value is 0.30000000000000004**

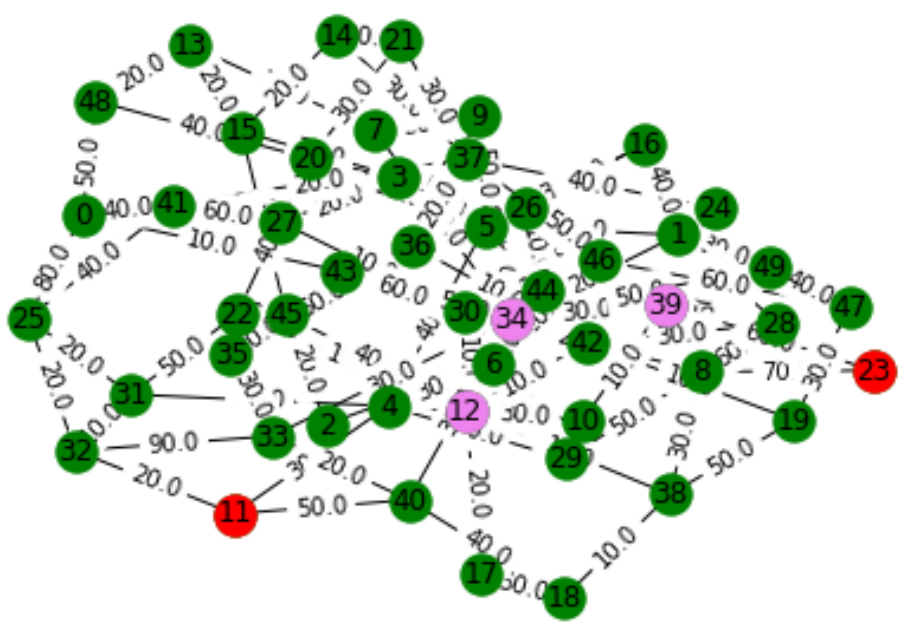

**Figure 16: Violet nodes are where IADS missile batteries have been placed and red nodes are unprotected location node numbers where IADS missile battery's interceptor missiles can't reach. First IADS missile battery is placed at location node 12, Second IADS missile battery is placed at location node 34 and Third IADS missile battery is placed at location node 39. First placed IADS missile battery's interceptor missile range is 80 km, second placed IADS missile battery's interceptor missile range is 70 km and third placed IADS missile battery's interceptor missile range is 20 km. Here, total unprotected asset value is 0.30000000000000004**

## Strategy_5:

(same as Strategy_3 and Strategy_4 for first four steps)

Step_1. Considering weighted edge network with asset values for placing IADS's batteries

Step_2. Computing network_algorithm_1, network_algorithm_2, network_algorithm_3, network_algorithm_4, network_algorithm_5 and network_algorithm_6

Step_3. Creating six sequences by ordering location node numbers as descending orders of the results of six network algorithms sequentially such as sequence_number_1 is the sequence of node numbers ordered as descending order of network_algorithm_1's measuring values on nodes.

Step_4. Estimating total unprotected asset values for placing IADS missile battery with maximum missile range on the place of first location node number from each sequence and choosing the lowest unprotected asset value corresponded node among six unprotected asset value corresponded nodes for placing first IADS with maximum missile range.

Step_5. After placing the first IADS missile battery with maximum missile range, we shall estimate total unprotected asset values (but for this strategy, unlike Strategy_3, we shall consider only those unprotected asset values which are not protected by previously placed IADS missile batteries) for placing next IADS missile battery with maximum missile range or with the rest of minimum missile ranges on the place of the next location node number where the distances from the first placed IADS missile battery is more than the first placed IADS missile battery's missile range in each sequence and choosing the lowest unprotected asset value corresponded location node among six unprotected asset value corresponded location nodes for placing the next IADS with maximum missile range or with the rest of minimum missile ranges.

Step_6. Repeating Step 5 for placing the next IADS missile battery with previous missile ranges or with the next longer missile range (which is less than previous missile range) until all IADS missile batteries are placed.

For example, One, two and three IADS missile batteries have been applied sequentially on that previously generated weighted edge location network where interceptor missile ranges are sequentially 80 km, 70 km and 20 km.

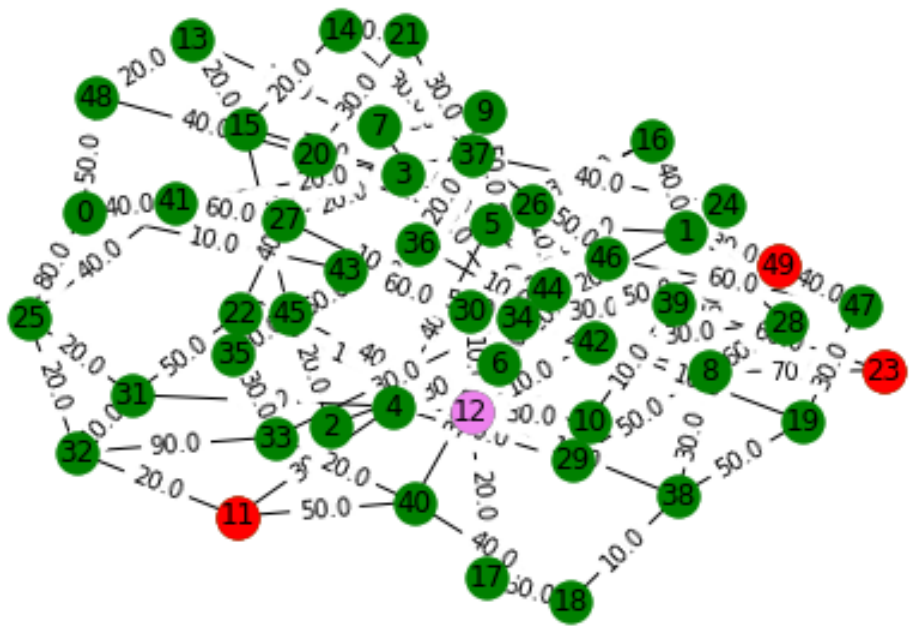

**Figure 17: Violet nodes are where IADS missile batteries have been placed and red nodes are unprotected location node numbers where IADS missile battery's interceptor missiles can't reach. One IADS missile battery is placed at location node number 12. Placed IADS missile battery's interceptor missile range is 80 km Here, total unprotected asset value is 0.5**

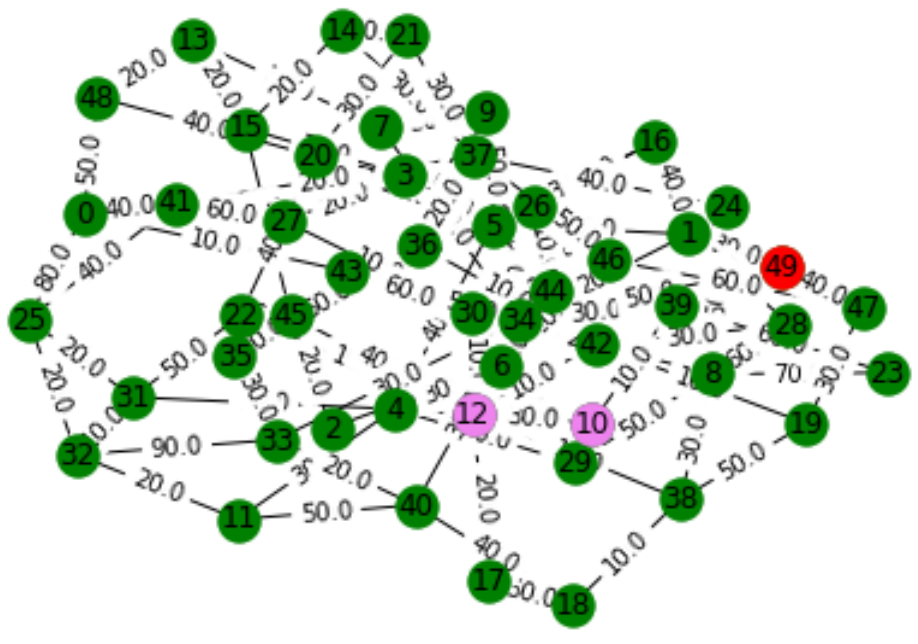

**Figure 18: Violet nodes are where IADS missile batteries have been placed and red nodes are unprotected location node numbers where IADS missile battery's interceptor missiles can't reach. First IADS missile battery is placed at location node number 12 and Second IADS missile battery is placed at**

**location node number 10. First placed IADS missile battery's interceptor missile range is 80 km and second placed IADS missile battery's interceptor missile range is 70 km. Here, total unprotected asset value is 0.2**

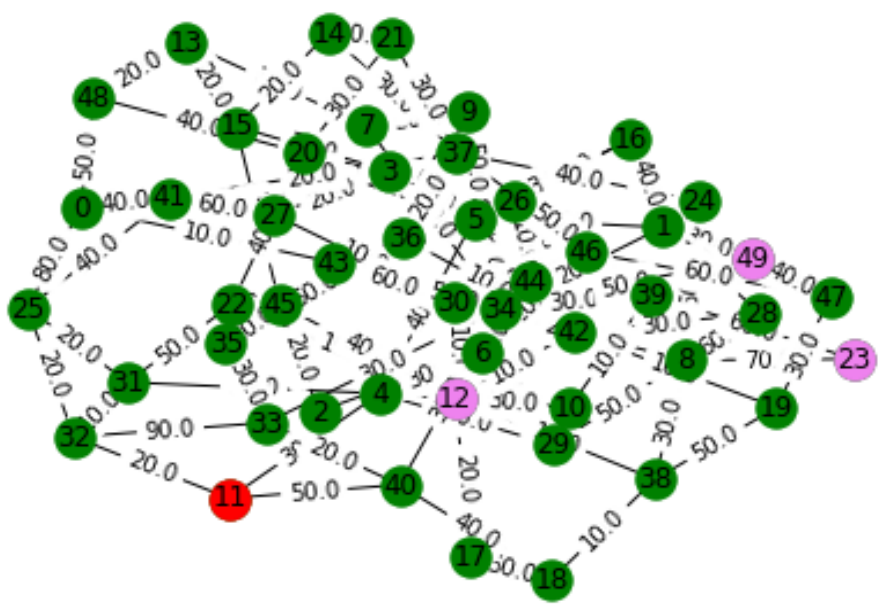

**Figure 19: Violet nodes are where IADS missile batteries have been placed and red nodes are unprotected location node numbers where IADS missile battery's interceptor missiles can't reach. First IADS missile battery is placed at location node 12, Second IADS missile battery is placed at location node 23 and third IADS missile battery is placed at location node 49. First placed IADS missile battery's interceptor missile range is 80 km, second placed IADS missile battery's interceptor missile range is 70 km and third placed IADS missile battery's interceptor missile range is 20 km. Here, total unprotected asset value is 0.1**

## Strategy_6:

(same as Strategy_3 and Strategy_4 for first four steps)

Step_1. Considering weighted edge network with asset values for placing IADS's batteries

Step_2. Computing network_algorithm_1, network_algorithm_2, network_algorithm_3, network_algorithm_4, network_algorithm_5 and network_algorithm_6

Step_3. Creating six sequences by ordering location node numbers as descending orders of the results of six network algorithms sequentially such as sequence_number_1 is the sequence of node numbers ordered as descending order of network_algorithm_1's measuring values on nodes.

Step_4. Estimating total unprotected asset values for placing IADS missile battery with maximum missile range on the place of first location node number from each sequence and choosing the lowest unprotected asset value corresponded node among six unprotected asset value corresponded nodes for placing first IADS with maximum missile range.

Step_5. After placing first IADS missile battery, we shall choose nodes from longest shortest distance path from first placed IADS missile battery and shortest path between previous shortest distance end node (end node of shortest distance path from first placed IADS missile battery) and first placed IADS missile battery location node.

Step_6. Including that, we shall choose nodes from shortest distance path between network diameter source node and first placed IADS_mssle_battery's location node and we shall also choose nodes from shortest distance path between network diameter target node and first placed IADS missile battery's location node.

Step_7. Creating sequence of chosen nodes without any order of ascending or descending and, from that sequence, we shall select optimal location nodes for placing the rest of IADS missile batteries based on less unprotected asset value cost as per rest of IADS missile batteries ranges.

For example, One, two and three IADS missile batteries have been applied sequentially on that previously generated weighted edge location network where interceptor missile ranges are sequentially 80 km, 70 km and 20 km.

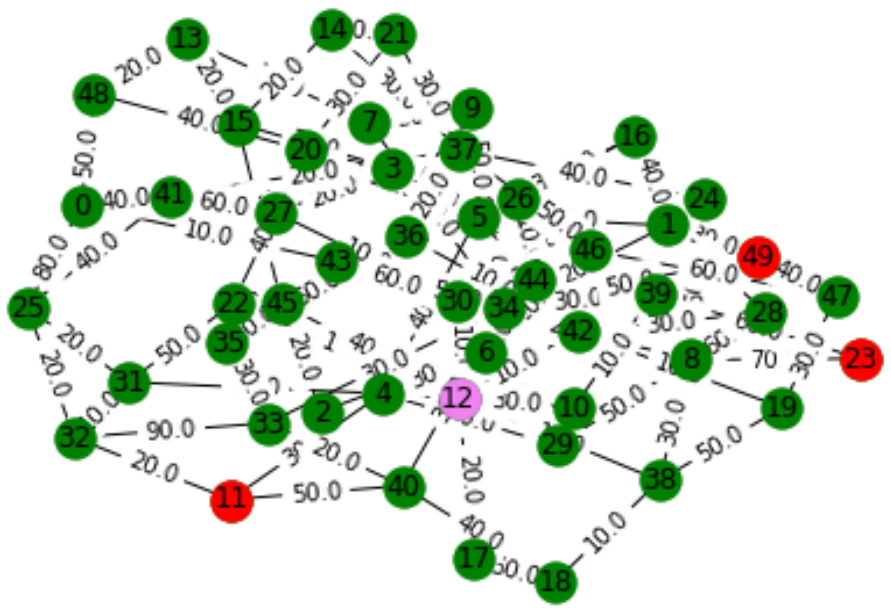

**Figure 20: Violet nodes are where IADS missile batteries have been placed and red nodes are unprotected location node numbers where IADS missile battery's interceptor missiles can't reach. One IADS missile battery is placed at location node number 12. Placed IADS missile battery's interceptor missile range is 80 km Here, total unprotected asset value is 0.5**

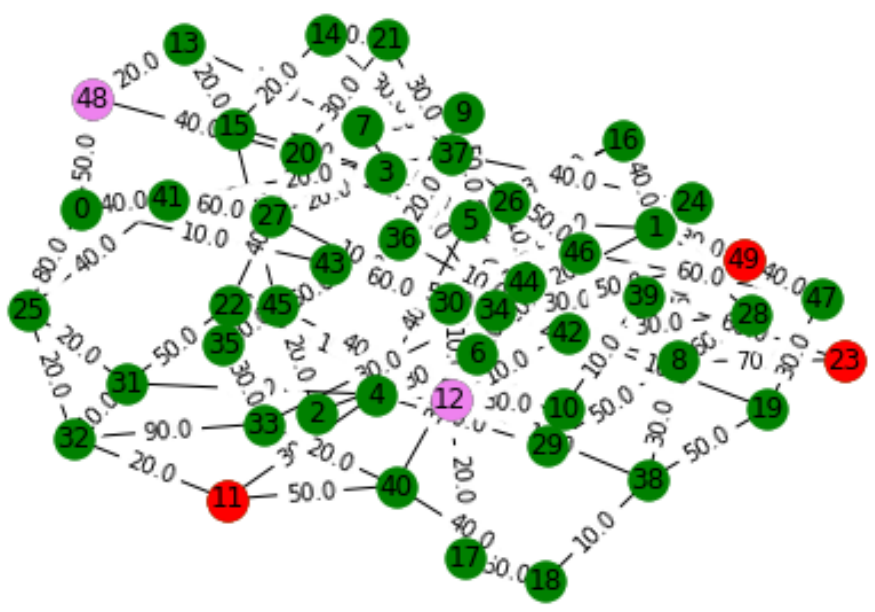

**Figure 21 : Violet nodes are where IADS missile batteries have been placed and red nodes are unprotected location node numbers where IADS missile battery's interceptor missiles can't reach. First IADS missile battery is placed at location node number 12 and Second IADS missile battery is placed at location node number 48. First placed IADS missile battery's interceptor missile range is 80 km and second placed IADS missile battery's interceptor missile range is 70 km. Here, total unprotected asset value is 0.5**

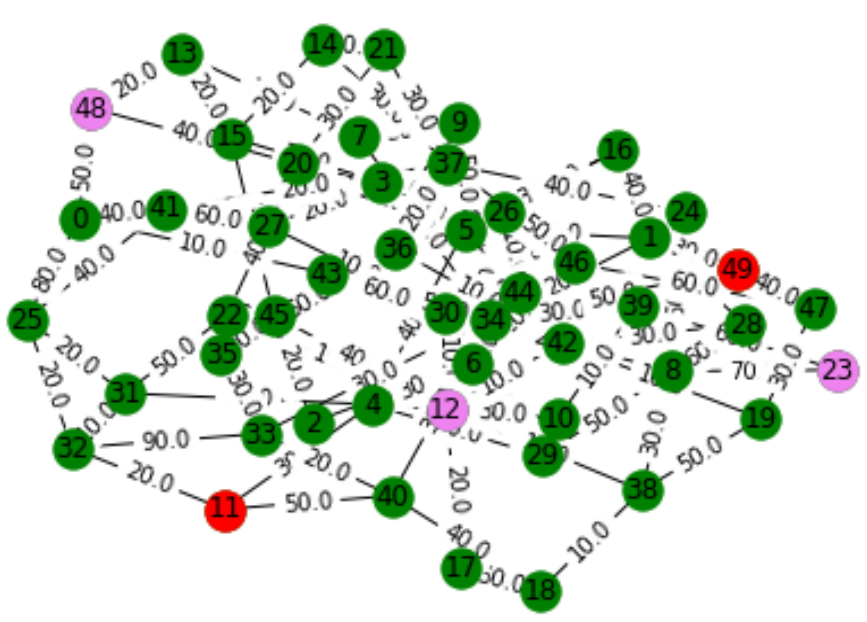

**Figure 22: Violet nodes are where IADS missile batteries have been placed and red nodes are unprotected location node numbers where IADS missile battery's interceptor missiles can't reach. First IADS missile battery is placed at location node 12, Second IADS missile battery is placed at location node 48 and third IADS missile battery is placed at location node 23. First placed IADS missile battery's interceptor missile range is 80 km, second placed IADS missile battery's interceptor missile range is 70 km and third placed IADS missile battery's interceptor missile range is 20 km. Here, total unprotected asset value is 0.30000000000000004**

## Strategy_7:

(same as Strategy_3 and Strategy_4 for first four steps)

Step_1. Considering weighted edge network with asset values for placing IADS's batteries

Step_2. Computing network_algorithm_1, network_algorithm_2, network_algorithm_3, network_algorithm_4, network_algorithm_5 and network_algorithm_6

Step_3. Creating six sequences by ordering location node numbers as descending orders of the results of six network algorithms sequentially such as sequence_number_1 is the sequence of node numbers ordered as descending order of network_algorithm_1's measuring values on nodes.

Step_4. Estimating total unprotected asset values for placing IADS missile battery with maximum missile range on the place of first location node number from each sequence and choosing the lowest unprotected asset value corresponded node among six unprotected asset value corresponded nodes for placing first IADS with maximum missile range.

Step_5. After placing first IADS missile battery, we shall choose nodes from longest shortest distance path from first placed IADS missile battery and shortest path between previous shortest distance end node (end node of shortest distance path from first placed IADS missile battery) and first placed IADS missile battery location node.

Step_6. Including that, we shall choose nodes from shortest distance path between network diameter source node and first placed IADS_mssle_battery's location node and we shall also choose nodes from shortest distance path between network diameter target node and first placed IADS missile battery's location node.

Step_7. Creating sequence of chosen nodes without any order of ascending or descending and from that sequence, creating another sequence from betweenness centrality subset of previously created subset.

Step_8. From the last created sequence, we shall select optimal location nodes for placing the rest of IADS missile batteries based on less unprotected asset value cost as per rest of IADS missile batteries ranges.

For example, One, two and three IADS missile batteries have been applied sequentially on that previously generated weighted edge location network where interceptor missile ranges are sequentially 80 km, 70 km and 20 km.

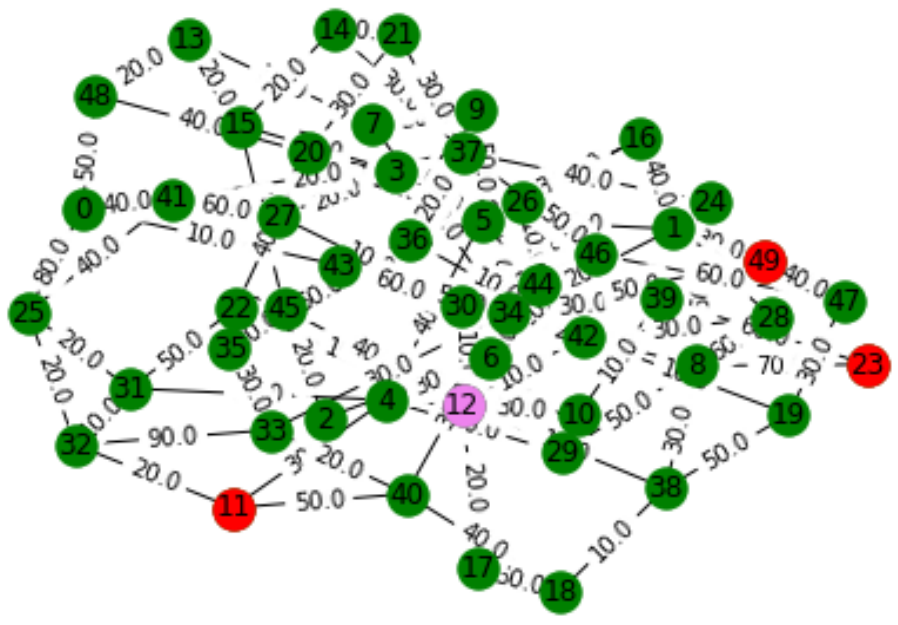

**Figure 23: Violet nodes are where IADS missile batteries have been placed and red nodes are unprotected location node numbers where IADS missile battery's interceptor missiles can't reach. One IADS missile battery is placed at location node number 12. Placed IADS missile battery's interceptor missile range is 80 km Here, total unprotected asset value is 0.5**

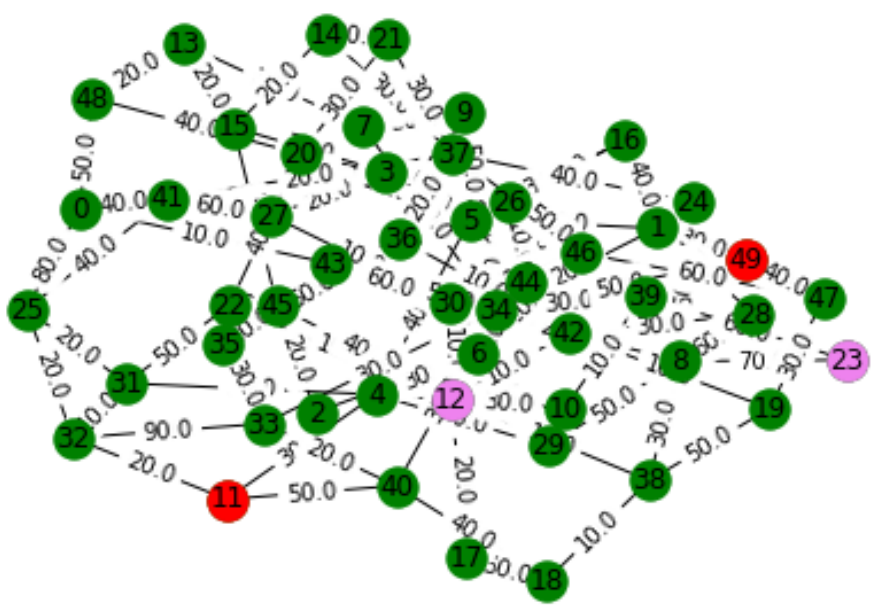

**Figure 24: Violet nodes are where IADS missile batteries have been placed and red nodes are unprotected location node numbers where IADS missile battery's interceptor missiles can't reach. First IADS missile battery is placed at location node number 12 and Second IADS missile battery is placed at**

**location node number 23. First placed IADS missile battery's interceptor missile range is 80 km and second placed IADS missile battery's interceptor missile range is 70 km. Here, total unprotected asset value is 0.30000000000000004**

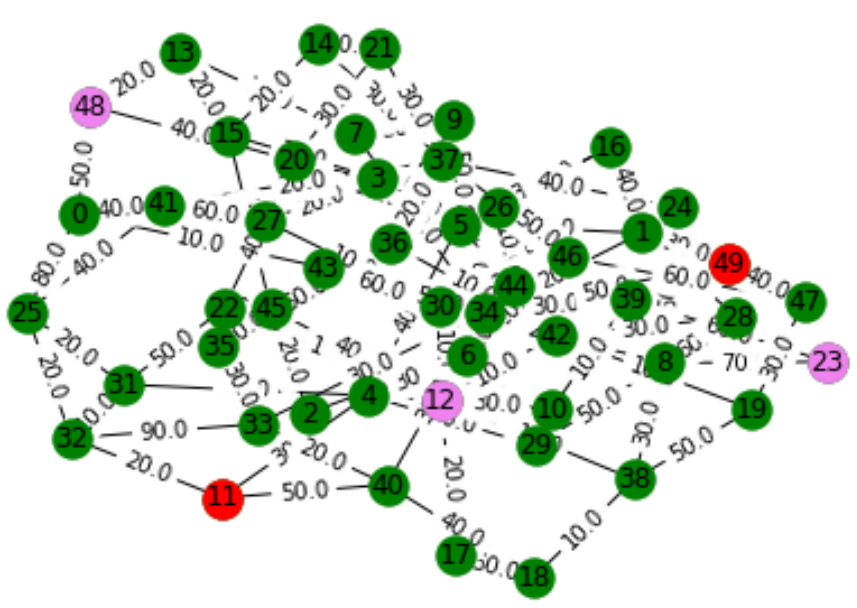

**Figure 25: Violet nodes are where IADS missile batteries have been placed and red nodes are unprotected location node numbers where IADS missile battery's interceptor missiles can't reach. First IADS missile battery is placed at location node 12, Second IADS missile battery is placed at location node 23 and third IADS missile battery is placed at location node 48. First placed IADS missile battery's interceptor missile range is 80 km, second placed IADS missile battery's interceptor missile range is 70 km and third placed IADS missile battery's interceptor missile range is 20 km. Here, total unprotected asset value is 0.30000000000000004**

## V. Computational Results

Table 1 has been created for results of two IADS missile batteries where interceptor missile battery ranges are sequentially 80 km and 70 km where worst case scenario's unprotected asset value percentages (including calculating interceptor missile's interceptor probabilities by considering only one IADS missile battery's one interceptor missile for one attacker missile) have been listed for few generated weighted edge networks and for each strategy.

| Strategy number | Generated weighted edge location network's small worldness | Generated weighted edge location network's node number | Generated weighted edge location network's Diameter | Generated weighted edge location network's Minimum edge length and Maximum edge length (Minimum, Maximum) | Generated weighted edge location network total asset value | IADS missile batteries Interceptor missile ranges in kilometer | IADS missile batteries interceptor missile's interceptor probability | After placing IADS missile batteries, total Unprotected asset value | After placing IADS missile batteries, total Unprotected asset value Percentage (Worst Case Scenario ) |
|---|---|---|---|---|---|---|---|---|---|
| **Strategy_1** | **≈0.39692** | **50** | **123.0 km** | **(20.5 km, 184.5 km)** | **≈14.9** | **80 km, 70 km** | **98%** | **1.6** | **≈12.52349%** |
| **Strategy_2** | **≈0.39692** | **50** | **123.0 km** | **(20.5 km, 184.5 km)** | **≈14.9** | **80 km, 70 km** | **98%** | **1.6** | **≈12.52349%** |
| **Strategy_3** | **≈0.39692** | **50** | **123.0 km** | **(20.5 km, 184.5 km)** | **≈14.9** | **80 km, 70 km** | **98%** | **0.6** | **≈5.94631%** |
| **Strategy_4** | **≈0.39692** | **50** | **123.0 km** | **(20.5 km, 184.5 km)** | **≈14.9** | **80 km, 70 km** | **98%** | **1.8** | **≈13.83893%** |
| **Strategy_5** | **≈0.39692** | **50** | **123.0 km** | **(20.5 km, 184.5 km)** | **≈14.9** | **80 km, 70 km** | **98%** | **0.2** | **≈3.315436%** |

| Strategy_6 | ≈0.39692 | 50 | 123.0 km | (20.5 km, 184.5 km) | ≈14.9 | 80 km, 70 km | 98% | 1.0 | ≈8.577181% |
|---|---|---|---|---|---|---|---|---|---|
| Strategy_7 | ≈0.39692 | 50 | 123.0 km | (20.5 km, 184.5 km) | ≈14.9 | 80 km, 70 km | 98% | 1.0 | ≈8.577181% |
| Strategy_1 | ≈0.582777 | 50 | 133.0 km | (19 km, 171.0 km) | ≈14.4 | 80 km, 70 km | 98% | 0.3 | ≈4.0416667% |
| Strategy_2 | ≈0.582777 | 50 | 133.0 km | (19 km, 171.0 km) | ≈14.4 | 80 km, 70 km | 98% | 0.3 | ≈4.0416667% |
| Strategy_3 | ≈0.582777 | 50 | 133.0 km | (19 km, 171.0 km) | ≈14.4 | 80 km, 70 km | 98% | 0.9 | ≈8.1250000% |
| Strategy_4 | ≈0.582777 | 50 | 133.0 km | (19 km, 171.0 km) | ≈14.4 | 80 km, 70 km | 98% | 0.4 | ≈4.7222222% |
| Strategy_5 | ≈0.582777 | 50 | 133.0 km | (19 km, 171.0 km) | ≈14.4 | 80 km, 70 km | 98% | 0.2 | ≈3.3611111% |
| Strategy_6 | ≈0.582777 | 50 | 133.0 km | (19 km, 171.0 km) | ≈14.4 | 80 km, 70 km | 98% | 0.3 | ≈4.0416667% |
| Strategy_7 | ≈0.582777 | 50 | 133.0 km | (19 km, 171.0 km) | ≈14.4 | 80 km, 70 km | 98% | 0.3 | ≈4.0416667% |
| Strategy_1 | ≈0.640091 | 50 | 200.0 km | (20.0 km, 160.0 km) | ≈14.3 | 80 km, 70 km | 98% | 4.1 | ≈30.097902% |
| Strategy_2 | ≈0.640091 | 50 | 200.0 km | (20.0 km, 160.0 km) | ≈14.3 | 80 km, 70 km | 98% | 4.5 | ≈32.839161% |
| Strategy_3 | ≈0.640091 | 50 | 200.0 km | (20.0 km, 160.0 km) | ≈14.3 | 80 km, 70 km | 98% | 3.3 | ≈24.615385% |
| Strategy_4 | ≈0.640091 | 50 | 200.0 km | (20.0 km, 160.0 km) | ≈14.3 | 80 km, 70 km | 98% | 1.3 | ≈10.90909% |
| Strategy_5 | ≈0.640091 | 50 | 200.0 km | (20.0 km, 160.0 km) | ≈14.3 | 80 km, 70 km | 98% | 1.3 | ≈10.90909% |
| Strategy_6 | ≈0.640091 | 50 | 200.0 km | (20.0 km, 160.0 km) | ≈14.3 | 80 km, 70 km | 98% | 2.7 | ≈20.503497% |
| Strategy_7 | ≈0.640091 | 50 | 200.0 km | (20.0 km, 160.0 km) | ≈14.3 | 80 km, 70 km | 98% | 2.4 | ≈18.447552% |
| Strategy_1 | ≈0.807039 | 50 | 710.0 km | (10 km, 70 km) | ≈15.4 | 80 km, 70 km | 98% | 10.5 | ≈68.181818% |
| Strategy_2 | ≈0.807039 | 50 | 710.0 km | (10 km, 70 km) | ≈15.4 | 80 km, 70 km | 98% | 12.5 | ≈81.545455% |
| Strategy_3 | ≈0.807039 | 50 | 710.0 km | (10 km, 70 km) | ≈15.4 | 80 km, 70 km | 98% | 7.2 | ≈47.818182% |
| Strategy_4 | ≈0.807039 | 50 | 710.0 km | (10 km, 70 km) | ≈15.4 | 80 km, 70 km | 98% | 10.6 | ≈69.454545% |
| Strategy_5 | ≈0.807039 | 50 | 710.0 km | (10 km, 70 km) | ≈15.4 | 80 km, 70 km | 98% | 6.6 | ≈44.00% |
| Strategy_6 | ≈0.807039 | 50 | 710.0 km | (10 km, | ≈15.4 | 80 km, | 98% | 6.6 | ≈44.00% |
| | | | | 70 km) | | 70 km | | | |
| **Strategy_7** | **≈0.807039** | **50** | **710.0 km** | **(10 km, 70 km)** | **≈15.4** | **80 km, 70 km** | **98%** | **6.6** | **≈44.00%** |
| **Strategy_1** | **≈1.145887** | **50** | **560.0 km** | **(10 km, 80 km)** | **≈14.1** | **80 km, 70 km** | **98%** | **9.2** | **≈65.943262%** |
| **Strategy_2** | **≈1.145887** | **50** | **560.0 km** | **(10 km, 80 km)** | **≈14.1** | **80 km, 70 km** | **98%** | **9.2** | **≈65.943262%** |
| **Strategy_3** | **≈1.145887** | **50** | **560.0 km** | **(10 km, 80 km)** | **≈14.1** | **80 km, 70 km** | **98%** | **8.4** | **≈60.382979%** |
| **Strategy_4** | **≈1.145887** | **50** | **560.0 km** | **(10 km, 80 km)** | **≈14.1** | **80 km, 70 km** | **98%** | **8.4** | **≈60.382979%** |
| **Strategy_5** | **≈1.145887** | **50** | **560.0 km** | **(10 km, 80 km)** | **≈14.1** | **80 km, 70 km** | **98%** | **7.9** | **≈56.907801%** |
| **Strategy_6** | **≈1.145887** | **50** | **560.0 km** | **(10 km, 80 km)** | **≈14.1** | **80 km, 70 km** | **98%** | **8.4** | **≈60.382979%** |
| **Strategy_7** | **≈1.145887** | **50** | **560.0 km** | **(10 km, 80 km)** | **≈14.1** | **80 km, 70 km** | **98%** | **7.9** | **≈56.907801%** |

**Table 1: Results for two IADS missile batteries where interceptor missile battery ranges are sequentially 80 km and 70 km where worst case scenario's unprotected asset value percentages (including calculating interceptor missile's interceptor probabilities by considering only one IADS missile battery's one interceptor missile for one attacker missile) have been listed for each of few generated weighted edge networks and for each strategy.**

Table 2 has been created for results for three IADS missile batteries where interceptor missile battery ranges are sequentially 110 km, 90 km and 80 km where worst case scenario's unprotected asset value percentages (including calculating interceptor missile's interceptor probabilities by considering only one IADS missile battery's one interceptor missile for one attacker missile) have been listed for few generated weighted edge networks and for each strategy.

| **Strategy number** | **Generated weighted edge location network's small worldness** | **Generated weighted edge location network's node number** | **Generated weighted edge location network's Diameter** | **Generated weighted edge location network's Minimum edge length and Maximum edge length (Minimum, Maximum)** | **Generated weighted edge location network total asset value** | **IADS missile batteries Interceptor missile ranges in kilometer** | **IADS missile batteries interceptor missile's interceptor probability** | **After placing IADS missile batteries, total Unprotected asset value** | **After placing IADS missile batteries, total Unprotected asset value Percentage (Worst Case Scenario )** |
|---|---|---|---|---|---|---|---|---|---|
| **Strategy_1** | **≈0.423134** | **50** | **210.0 km** | **(35 km, 315 km)** | **13.8** | **110 km, 90 km, 80 km** | **98%** | **0.9** | **≈8.39130%** |
| **Strategy_2** | **≈0.423134** | **50** | **210.0 km** | **(35 km, 315 km)** | **13.8** | **110 km, 90 km, 80 km** | **98%** | **1.0** | **≈9.101449%** |
| **Strategy_3** | **≈0.423134** | **50** | **210.0 km** | **(35 km, 315 km)** | **13.8** | **110 km, 90 km, 80 km** | **98%** | **0.6** | **≈6.26087%** |
| **Strategy_4** | **≈0.423134** | **50** | **210.0 km** | **(35 km,** | **13.8** | **110 km,** | **98%** | **0.1** | **≈2.71014%** |
| | | | | 315 km) | | 90 km, 80 km | | | |
| Strategy_5 | ≈0.423134 | 50 | 210.0 km | (35 km, 315 km) | 13.8 | 110 km, 90 km, 80 km | 98% | 0.0 | 2.0% |
| Strategy_6 | ≈0.423134 | 50 | 210.0 km | (35 km, 315 km) | 13.8 | 110 km, 90 km, 80 km | 98% | 0.4 | ≈4.84058% |
| Strategy_7 | ≈0.423134 | 50 | 210.0 km | (35 km, 315 km) | 13.8 | 110 km, 90 km, 80 km | 98% | 0.4 | ≈4.84058% |
| Strategy_1 | ≈0.564176 | 50 | 270.0 km | (18 km, 144 km) | 15.0 | 110 km, 90 km, 80 km | 98% | 4.7 | ≈32.706667% |
| Strategy_2 | ≈0.564176 | 50 | 270.0 km | (18 km, 144 km) | 15.0 | 110 km, 90 km, 80 km | 98% | 5.5 | ≈37.933333% |
| Strategy_3 | ≈0.564176 | 50 | 270.0 km | (18 km, 144 km) | 15.0 | 110 km, 90 km, 80 km | 98% | 4.7 | ≈32.706667% |
| Strategy_4 | ≈0.564176 | 50 | 270.0 km | (18 km, 144 km) | 15.0 | 110 km, 90 km, 80 km | 98% | 2.1 | ≈15.72% |
| Strategy_5 | ≈0.564176 | 50 | 270.0 km | (18 km, 144 km) | 15.0 | 110 km, 90 km, 80 km | 98% | 2.1 | ≈15.72% |
| Strategy_6 | ≈0.564176 | 50 | 270.0 km | (18 km, 144 km) | 15.0 | 110 km, 90 km, 80 km | 98% | 4.5 | ≈31.4% |
| Strategy_7 | ≈0.564176 | 50 | 270.0 km | (18 km, 144 km) | 15.0 | 110 km, 90 km, 80 km | 98% | 4.5 | ≈31.4% |
| Strategy_1 | ≈0.640606 | 50 | 200.0 km | (25 km, 225 km) | 15.7 | 110 km, 90 km, 80 km | 98% | 0.3 | ≈3.872611% |
| Strategy_2 | ≈0.640606 | 50 | 200.0 km | (25 km, 225 km) | 15.7 | 110 km, 90 km, 80 km | 98% | 1.1 | ≈8.866242% |
| Strategy_3 | ≈0.640606 | 50 | 200.0 km | (25 km, 225 km) | 15.7 | 110 km, 90 km, 80 km | 98% | 2.4 | ≈16.980891% |
| Strategy_4 | ≈0.640606 | 50 | 200.0 km | (25 km, 225 km) | 15.7 | 110 km, 90 km, 80 km | 98% | 1.1 | ≈8.866242% |
| Strategy_5 | ≈0.640606 | 50 | 200.0 km | (25 km, 225 km) | 15.7 | 110 km, 90 km, 80 km | 98% | 0.1 | ≈2.6242038% |
| Strategy_6 | ≈0.640606 | 50 | 200.0 km | (25 km, 225 km) | 15.7 | 110 km, 90 km, | 98% | 1.0 | ≈8.2420382% |

| | | | | | | 80 km | | | |
|---|---|---|---|---|---|---|---|---|---|
| **Strategy_7** | **≈0.640606** | **50** | **200.0 km** | **(25 km, 225 km)** | **15.7** | **110 km, 90 km, 80 km** | **98%** | **1.0** | **≈8.2420382%** |
| **Strategy_1** | **≈0.829442** | **50** | **345.0 km** | **(23 km, 184 km)** | **15.4** | **110 km, 90 km, 80 km** | **98%** | **5.1** | **≈34.454545%** |
| **Strategy_2** | **≈0.829442** | **50** | **345.0 km** | **(23 km, 184 km)** | **15.4** | **110 km, 90 km, 80 km** | **98%** | **8.9** | **≈58.636363%** |
| **Strategy_3** | **≈0.829442** | **50** | **345.0 km** | **(23 km, 184 km)** | **15.4** | **110 km, 90 km, 80 km** | **98%** | **8.0** | **≈52.909090%** |
| **Strategy_4** | **≈0.829442** | **50** | **345.0 km** | **(23 km, 184 km)** | **15.4** | **110 km, 90 km, 80 km** | **98%** | **4.1** | **≈28.090909%** |
| **Strategy_5** | **≈0.829442** | **50** | **345.0 km** | **(23 km, 184 km)** | **15.4** | **110 km, 90 km, 80 km** | **98%** | **3.5** | **≈24.272727%** |
| **Strategy_6** | **≈0.829442** | **50** | **345.0 km** | **(23 km, 184 km)** | **15.4** | **110 km, 90 km, 80 km** | **98%** | **5.8** | **≈38.90909%** |
| **Strategy_7** | **≈0.829442** | **50** | **345.0 km** | **(23 km, 184 km)** | **15.4** | **110 km, 90 km, 80 km** | **98%** | **6.2** | **≈41.454545%** |
| **Strategy_1** | **≈1.16390** | **50** | **960.0 km** | **(20 km, 160 km)** | **14.2** | **110 km, 90 km, 80 km** | **98%** | **10.5** | **≈74.464789%** |
| **Strategy_2** | **≈1.16390** | **50** | **960.0 km** | **(20 km, 160 km)** | **14.2** | **110 km, 90 km, 80 km** | **98%** | **10.8** | **≈76.535211%** |
| **Strategy_3** | **≈1.16390** | **50** | **960.0 km** | **(20 km, 160 km)** | **14.2** | **110 km, 90 km, 80 km** | **98%** | **10.8** | **≈76.535211%** |
| **Strategy_4** | **≈1.16390** | **50** | **960.0 km** | **(20 km, 160 km)** | **14.2** | **110 km, 90 km, 80 km** | **98%** | **10.3** | **≈73.084507%** |
| **Strategy_5** | **≈1.16390** | **50** | **960.0 km** | **(20 km, 160 km)** | **14.2** | **110 km, 90 km, 80 km** | **98%** | **9.4** | **≈66.873239%** |
| **Strategy_6** | **≈1.16390** | **50** | **960.0 km** | **(20 km, 160 km)** | **14.2** | **110 km, 90 km, 80 km** | **98%** | **10.2** | **≈72.394366%** |
| **Strategy_7** | **≈1.16390** | **50** | **960.0 km** | **(20 km, 160 km)** | **14.2** | **110 km, 90 km, 80 km** | **98%** | **11.2** | **≈79.295775%** |

**Table 2: Results for three IADS missile batteries where interceptor missile battery ranges are sequentially 110 km, 90 km and 80 km where worst case scenario's unprotected asset value percentages (including calculating interceptor missile's interceptor probabilities by considering only one IADS missile battery's one interceptor missile for one attacker missile) have been listed for each of few generated weighted edge networks and for each strategy.**

Table 3 has been created for results for four IADS missile batteries where interceptor missile battery ranges are sequentially 120 km, 110 km, 90 km and 80 km where worst case scenario's unprotected asset value percentages (including calculating interceptor missile's interceptor probabilities by considering only one IADS missile battery's one interceptor missile for one attacker missile) have been listed for few generated weighted edge networks and for each strategy.

| **Strategy number** | **Generated weighted edge location network's small worldness** | **Generated weighted edge location network's node number** | **Generated weighted edge location network's Diameter** | **Generated weighted edge location network's Minimum edge length and Maximum edge length (Minimum, Maximum)** | **Generated weighted edge location network total asset value** | **IADS missile batteries Interceptor missile ranges in kilometer** | **IADS missile batteries interceptor missile's interceptor probability** | **After placing IADS missile batteries, total Unprotected asset value** | **After placing IADS missile batteries, total Unprotected asset value Percentage (Worst Case Scenario )** |
|---|---|---|---|---|---|---|---|---|---|
| **Strategy_1** | **≈0.25351** | **50** | **270.0 km** | **(45 km, 405 km)** | **14.1** | **120 km, 110 km, 90 km, 80 km** | **98%** | **2.4** | **≈18.68085%** |
| **Strategy_2** | **≈0.25351** | **50** | **270.0 km** | **(45 km, 405 km)** | **14.1** | **120 km, 110 km, 90 km, 80 km** | **98%** | **3.8** | **≈28.4113%** |
| **Strategy_3** | **≈0.25351** | **50** | **270.0 km** | **(45 km, 405 km)** | **14.1** | **120 km, 110 km, 90 km, 80 km** | **98%** | **1.7** | **≈13.8156%** |
| **Strategy_4** | **≈0.25351** | **50** | **270.0 km** | **(45 km, 405 km)** | **14.1** | **120 km, 110 km, 90 km, 80 km** | **98%** | **1.4** | **≈11.730%** |
| **Strategy_5** | **≈0.25351** | **50** | **270.0 km** | **(45 km, 405 km)** | **14.1** | **120 km, 110 km, 90 km, 80 km** | **98%** | **0.3** | **≈4.0851%** |
| **Strategy_6** | **≈0.25351** | **50** | **270.0 km** | **(45 km, 405 km)** | **14.1** | **120 km, 110 km, 90 km, 80 km** | **98%** | **1.9** | **≈15.20567%** |
| **Strategy_7** | **≈0.25351** | **50** | **270.0 km** | **(45 km, 405 km)** | **14.1** | **120 km, 110 km, 90 km, 80 km** | **98%** | **1.6** | **≈13.120567%** |
| **Strategy_1** | **≈0.405311** | **50** | **360.0 km** | **(60 km, 540 km)** | **13.2** | **120 km, 110 km, 90 km, 80 km** | **98%** | **4.9** | **≈38.378788%** |
| **Strategy_2** | **≈0.405311** | **50** | **360.0 km** | **(60 km, 540 km)** | **13.2** | **120 km, 110 km, 90 km,** | **98%** | **6.9** | **≈53.227272%** |

| | | | | | | 80 km | | | |
|---|---|---|---|---|---|---|---|---|---|
| Strategy_3 | ≈0.405311 | 50 | 360.0 km | (60 km, 540 km) | 13.2 | 120 km, 110 km, 90 km, 80 km | 98% | 6.2 | ≈48.030303% |
| Strategy_4 | ≈0.405311 | 50 | 360.0 km | (60 km, 540 km) | 13.2 | 120 km, 110 km, 90 km, 80 km | 98% | 5.7 | ≈44.318181% |
| Strategy_5 | ≈0.405311 | 50 | 360.0 km | (60 km, 540 km) | 13.2 | 120 km, 110 km, 90 km, 80 km | 98% | 5.3 | ≈41.34848% |
| Strategy_6 | ≈0.405311 | 50 | 360.0 km | (60 km, 540 km) | 13.2 | 120 km, 110 km, 90 km, 80 km | 98% | 7.7 | ≈59.2% |
| Strategy_7 | ≈0.405311 | 50 | 360.0 km | (60 km, 540 km) | 13.2 | 120 km, 110 km, 90 km, 80 km | 98% | 5.6 | ≈43.575757% |
| Strategy_1 | ≈0.652449 | 50 | 375.0 km | (25 km, 200 km) | 16.2 | 120 km, 110 km, 90 km, 80 km | 98% | 5.2 | ≈33.45679% |
| Strategy_2 | ≈0.652449 | 50 | 375.0 km | (25 km, 200 km) | 16.2 | 120 km, 110 km, 90 km, 80 km | 98% | 4.6 | ≈29.82716% |
| Strategy_3 | ≈0.652449 | 50 | 375.0 km | (25 km, 200 km) | 16.2 | 120 km, 110 km, 90 km, 80 km | 98% | 3.0 | ≈20.14815% |
| Strategy_4 | ≈0.652449 | 50 | 375.0 km | (25 km, 200 km) | 16.2 | 120 km, 110 km, 90 km, 80 km | 98% | 3.6 | ≈23.7778% |
| Strategy_5 | ≈0.652449 | 50 | 375.0 km | (25 km, 200 km) | 16.2 | 120 km, 110 km, 90 km, 80 km | 98% | 1.5 | ≈11.074074% |
| Strategy_6 | ≈0.652449 | 50 | 375.0 km | (25 km, 200 km) | 16.2 | 120 km, 110 km, 90 km, 80 km | 98% | 5.1 | ≈32.85185% |
| Strategy_7 | ≈0.652449 | 50 | 375.0 km | (25 km, 200 km) | 16.2 | 120 km, 110 km, 90 km, 80 km | 98% | 4.7 | ≈30.4321% |
| Strategy_1 | ≈0.934170 | 50 | 697.0 km | (17 km, 119 km) | 14.6 | 120 km, 110 km, 90 km, 80 km | 98% | 6.7 | ≈46.9726% |

| Strategy_2 | ≈0.934170 | 50 | 697.0 km | (17 km, 119 km) | 14.6 | 120 km, 110 km, 90 km, 80 km | 98% | 8.6 | ≈59.7260% |
|---|---|---|---|---|---|---|---|---|---|
| Strategy_3 | ≈0.934170 | 50 | 697.0 km | (17 km, 119 km) | 14.6 | 120 km, 110 km, 90 km, 80 km | 98% | 6.8 | ≈47.64383% |
| Strategy_4 | ≈0.934170 | 50 | 697.0 km | (17 km, 119 km) | 14.6 | 120 km, 110 km, 90 km, 80 km | 98% | 5.7 | ≈40.26027% |
| Strategy_5 | ≈0.934170 | 50 | 697.0 km | (17 km, 119 km) | 14.6 | 120 km, 110 km, 90 km, 80 km | 98% | 3.0 | ≈22.136986% |
| Strategy_6 | ≈0.934170 | 50 | 697.0 km | (17 km, 119 km) | 14.6 | 120 km, 110 km, 90 km, 80 km | 98% | 3.1 | ≈22.8082% |
| Strategy_7 | ≈0.934170 | 50 | 697.0 km | (17 km, 119 km) | 14.6 | 120 km, 110 km, 90 km, 80 km | 98% | 4.0 | ≈28.849315% |
| Strategy_1 | ≈1.11291 | 50 | 960.0 km | (15 km, 135 km) | 16.1 | 120 km, 110 km, 90 km, 80 km | 98% | 9.7 | ≈61.043478% |
| Strategy_2 | ≈1.11291 | 50 | 960.0 km | (15 km, 135 km) | 16.1 | 120 km, 110 km, 90 km, 80 km | 98% | 10.3 | ≈64.695652% |
| Strategy_3 | ≈1.11291 | 50 | 960.0 km | (15 km, 135 km) | 16.1 | 120 km, 110 km, 90 km, 80 km | 98% | 11.4 | ≈71.39130% |
| Strategy_4 | ≈1.11291 | 50 | 960.0 km | (15 km, 135 km) | 16.1 | 120 km, 110 km, 90 km, 80 km | 98% | 11.8 | ≈73.826089% |
| Strategy_5 | ≈1.11291 | 50 | 960.0 km | (15 km, 135 km) | 16.1 | 120 km, 110 km, 90 km, 80 km | 98% | 7.0 | ≈44.608696% |
| Strategy_6 | ≈1.11291 | 50 | 960.0 km | (15 km, 135 km) | 16.1 | 120 km, 110 km, 90 km, 80 km | 98% | 9.7 | ≈61.043478% |
| Strategy_7 | ≈1.11291 | 50 | 960.0 km | (15 km, 135 km) | 16.1 | 120 km, 110 km, 90 km, 80 km | 98% | 10.0 | ≈62.869565% |

**Table 3: Results for four IADS missile batteries where interceptor missile battery ranges are sequentially 120 km, 110 km, 90 km and 80 km where worst case scenario's unprotected asset value percentages (including calculating interceptor missile's interceptor**

**probabilities by considering only one IADS missile battery's one interceptor missile for one attacker missile) have been listed for each of few generated weighted edge networks and for each strategy.**

Table 4 has been created for results for five IADS missile batteries where interceptor missile battery ranges are sequentially 200 km, 120 km, 110 km, 90 km and 80 km where worst case scenario's unprotected asset value percentages (including calculating interceptor missile's interceptor probabilities by considering only one IADS missile battery's one interceptor missile for one attacker missile) have been listed for few generated weighted edge networks and for each strategy.

| Strategy number | Generated weighted edge location network's small worldness | Generated weighted edge location network's node number | Generated weighted edge location network's Diameter | Generated weighted edge location network's Minimum edge length and Maximum edge length (Minimum, Maximum) | Generated weighted edge location network total asset value | IADS missile batteries Interceptor missile ranges in kilometer | IADS missile batteries interceptor missile's interceptor probability | After placing IADS missile batteries, total Unprotected asset value | After placing IADS missile batteries, total Unprotected asset value Percentage (Worst Case Scenario ) |
|---|---|---|---|---|---|---|---|---|---|
| Strategy_1 | ≈0.344096 | 50 | 480.0 km | (80.0 km, 640.0 km) | 16.2 | 200 km, 120 km, 110 km, 90 km, 80 km | 98% | 3.3 | ≈21.96296% |
| Strategy_2 | ≈0.344096 | 50 | 480.0 km | (80.0 km, 640.0 km) | 16.2 | 200 km, 120 km, 110 km, 90 km, 80 km | 98% | 3.8 | ≈24.987654% |
| Strategy_3 | ≈0.344096 | 50 | 480.0 km | (80.0 km, 640.0 km) | 16.2 | 200 km, 120 km, 110 km, 90 km, 80 km | 98% | 4.5 | ≈29.2222% |
| Strategy_4 | ≈0.344096 | 50 | 480.0 km | (80.0 km, 640.0 km) | 16.2 | 200 km, 120 km, 110 km, 90 km, 80 km | 98% | 3.8 | ≈24.98765% |
| Strategy_5 | ≈0.344096 | 50 | 480.0 km | (80.0 km, 640.0 km) | 16.2 | 200 km, 120 km, 110 km, 90 km, 80 km | 98% | 2.3 | ≈15.91358% |
| Strategy_6 | ≈0.344096 | 50 | 480.0 km | (80.0 km, 640.0 km) | 16.2 | 200 km, 120 km, 110 km, 90 km, 80 km | 98% | 4.0 | ≈26.19753% |
| Strategy_7 | ≈0.344096 | 50 | 480.0 km | (80.0 km, 640.0 km) | 16.2 | 200 km, 120 km, 110 km, 90 km, 80 km | 98% | 3.7 | ≈24.382716% |

| Strategy_1 | ≈0.431319 | 50 | 375.0 km | (75.0 km, 675.0 km) | 14.7 | 200 km, 120 km, 110 km, 90 km, 80 km | 98% | 8.6 | ≈59.333% |
|---|---|---|---|---|---|---|---|---|---|
| Strategy_2 | ≈0.431319 | 50 | 375.0 km | (75.0 km, 675.0 km) | 14.7 | 200 km, 120 km, 110 km, 90 km, 80 km | 98% | 10.5 | 72.0% |
| Strategy_3 | ≈0.431319 | 50 | 375.0 km | (75.0 km, 675.0 km) | 14.7 | 200 km, 120 km, 110 km, 90 km, 80 km | 98% | 5.3 | ≈37.3333% |
| Strategy_4 | ≈0.431319 | 50 | 375.0 km | (75.0 km, 675.0 km) | 14.7 | 200 km, 120 km, 110 km, 90 km, 80 km | 98% | 4.4 | ≈31.333% |
| Strategy_5 | ≈0.431319 | 50 | 375.0 km | (75.0 km, 675.0 km) | 14.7 | 200 km, 120 km, 110 km, 90 km, 80 km | 98% | 2.6 | ≈19.33333% |
| Strategy_6 | ≈0.431319 | 50 | 375.0 km | (75.0 km, 675.0 km) | 14.7 | 200 km, 120 km, 110 km, 90 km, 80 km | 98% | 7.3 | ≈50.66667% |
| Strategy_7 | ≈0.431319 | 50 | 375.0 km | (75.0 km, 675.0 km) | 14.7 | 200 km, 120 km, 110 km, 90 km, 80 km | 98% | 4.6 | ≈32.6666% |
| Strategy_1 | ≈0.652398 | 50 | 550.0 km | (50.0 km, 400.0 km) | 15.3 | 200 km, 120 km, 110 km, 90 km, 80 km | 98% | 5.5 | ≈37.228758% |
| Strategy_2 | ≈0.652398 | 50 | 550.0 km | (50.0 km, 400.0 km) | 15.3 | 200 km, 120 km, 110 km, 90 km, 80 km | 98% | 6.9 | ≈46.196078% |
| Strategy_3 | ≈0.652398 | 50 | 550.0 km | (50.0 km, 400.0 km) | 15.3 | 200 km, 120 km, 110 km, 90 km, 80 km | 98% | 6.4 | ≈42.993464% |
| Strategy_4 | ≈0.652398 | 50 | 550.0 km | (50.0 km, 400.0 km) | 15.3 | 200 km, 120 km, 110 km, 90 km, 80 km | 98% | 4.7 | ≈32.104575% |

| Strategy_5 | ≈0.652398 | 50 | 550.0 km | (50.0 km, 400.0 km) | 15.3 | 200 km,<br>120 km,<br>110 km,<br>90 km,<br>80 km | 98% | 2.6 | ≈18.65359% |
|---|---|---|---|---|---|---|---|---|---|
| Strategy_6 | ≈0.652398 | 50 | 550.0 km | (50.0 km, 400.0 km) | 15.3 | 200 km,<br>120 km,<br>110 km,<br>90 km,<br>80 km | 98% | 6.8 | ≈45.555556% |
| Strategy_7 | ≈0.652398 | 50 | 550.0 km | (50.0 km, 400.0 km) | 15.3 | 200 km,<br>120 km,<br>110 km,<br>90 km,<br>80 km | 98% | 5.1 | ≈34.666667% |
| Strategy_1 | ≈0.880917 | 50 | 600.0 km | (40.0 km, 360.0 km) | 14.7 | 200 km,<br>120 km,<br>110 km,<br>90 km,<br>80 km | 98% | 4.9 | ≈34.666667% |
| Strategy_2 | ≈0.880917 | 50 | 600.0 km | (40.0 km, 360.0 km) | 14.7 | 200 km,<br>120 km,<br>110 km,<br>90 km,<br>80 km | 98% | 6.7 | ≈46.666667% |
| Strategy_3 | ≈0.880917 | 50 | 600.0 km | (40.0 km, 360.0 km) | 14.7 | 200 km,<br>120 km,<br>110 km,<br>90 km,<br>80 km | 98% | 5.5 | ≈38.666667% |
| Strategy_4 | ≈0.880917 | 50 | 600.0 km | (40.0 km, 360.0 km) | 14.7 | 200 km,<br>120 km,<br>110 km,<br>90 km,<br>80 km | 98% | 4.4 | ≈31.33333% |
| Strategy_5 | ≈0.880917 | 50 | 600.0 km | (40.0 km, 360.0 km) | 14.7 | 200 km,<br>120 km,<br>110 km,<br>90 km,<br>80 km | 98% | 1.8 | 14.0% |
| Strategy_6 | ≈0.880917 | 50 | 600.0 km | (40.0 km, 360.0 km) | 14.7 | 200 km,<br>120 km,<br>110 km,<br>90 km,<br>80 km | 98% | 6.5 | ≈45.333333% |
| Strategy_7 | ≈0.880917 | 50 | 600.0 km | (40.0 km, 360.0 km) | 14.7 | 200 km,<br>120 km,<br>110 km,<br>90 km,<br>80 km | 98% | 5.7 | ≈40.00% |
| Strategy_1 | ≈1.12557 | 50 | 1425.0 km | (25.0 km, 250.0 km) | 14.0 | 200 km,<br>120 km,<br>110 km,<br>90 km,<br>80 km | 98% | 6.5 | ≈47.5% |
| **Strategy_2** | **≈1.12557** | **50** | **1425.0 km** | **(25.0 km, 250.0 km)** | **14.0** | **200 km,<br>120 km,<br>110 km,<br>90 km,<br>80 km** | **98%** | **7.9** | **≈57.3%** |
| **Strategy_3** | **≈1.12557** | **50** | **1425.0 km** | **(25.0 km, 250.0 km)** | **14.0** | **200 km,<br>120 km,<br>110 km,<br>90 km,<br>80 km** | **98%** | **7.5** | **≈54.5%** |
| **Strategy_4** | **≈1.12557** | **50** | **1425.0 km** | **(25.0 km, 250.0 km)** | **14.0** | **200 km,<br>120 km,<br>110 km,<br>90 km,<br>80 km** | **98%** | **7.1** | **≈51.7%** |
| **Strategy_5** | **≈1.12557** | **50** | **1425.0 km** | **(25.0 km, 250.0 km)** | **14.0** | **200 km,<br>120 km,<br>110 km,<br>90 km,<br>80 km** | **98%** | **4.1** | **≈30.7%** |
| **Strategy_6** | **≈1.12557** | **50** | **1425.0 km** | **(25.0 km, 250.0 km)** | **14.0** | **200 km,<br>120 km,<br>110 km,<br>90 km,<br>80 km** | **98%** | **6.6** | **≈48.2%** |
| **Strategy_7** | **≈1.12557** | **50** | **1425.0 km** | **(25.0 km, 250.0 km)** | **14.0** | **200 km,<br>120 km,<br>110 km,<br>90 km,<br>80 km** | **98%** | **6.9** | **≈50.3%** |

**Table 4: Results for five IADS missile batteries where interceptor missile battery ranges are sequentially 200km, 120 km, 110 km, 90 km and 80 km where worst case scenario's unprotected asset value percentages (including calculating interceptor missile's interceptor probabilities by considering only one IADS missile battery's one interceptor missile for one attacker missile) have been listed for each of few generated weighted edge networks and for each strategy.**

By aggregating each optimal solution for each generated network, we found Table 5.

| **Strategy number** | **Generated weighted edge location network's small worldness** | **Generated weighted edge location network's node number** | **Generated weighted edge location network's Diameter** | **Generated weighted edge location network's Minimum edge length and Maximum edge length (Minimum, Maximum)** | **Generated weighted edge location network total asset value** | **IADS missile batteries Interceptor missile ranges in kilometer** | **IADS missile batteries interceptor missile's interceptor probability** | **After placing IADS missile batteries, total Unprotected asset value** | **After placing IADS missile batteries, total Unprotected asset value Percentage (Worst Case Scenario )** |
|---|---|---|---|---|---|---|---|---|---|
| **Strategy_5** | **≈0.39692** | **50** | **123.0 km** | **(20.5 km, 184.5 km)** | **≈14.9** | **80 km,<br>70 km** | **98%** | **0.2** | **≈3.315436%** |
| **Strategy_5** | **≈0.582777** | **50** | **133.0 km** | **(19 km, 171.0 km)** | **≈14.4** | **80 km,<br>70 km** | **98%** | **0.2** | **≈3.3611111%** |
| **Strategy_5** | **≈0.640091** | **50** | **200.0 km** | **(20.0 km, 160.0 km)** | **≈14.3** | **80 km,<br>70 km** | **98%** | **1.3** | **≈10.90909%** |

| Strategy_5<br>Or,<br>Strategy_6<br>Or,<br>Strategy_7 | ≈0.807039 | 50 | 710.0 km | (10 km,<br>70 km) | ≈15.4 | 80 km,<br>70 km | 98% | 6.6 | ≈44.00% |
|---|---|---|---|---|---|---|---|---|---|
| Strategy_5 | ≈1.145887 | 50 | 560.0 km | (10 km,<br>80 km) | ≈14.1 | 80 km,<br>70 km | 98% | 7.9 | ≈56.907801% |
| Strategy_5 | ≈0.423134 | 50 | 210.0 km | (35 km,<br>315 km) | 13.8 | 110 km,<br>90 km,<br>80 km | 98% | 0.0 | 2.0% |
| Strategy_4<br>Or,<br>Strategy_5 | ≈0.564176 | 50 | 270.0 km | (18 km,<br>144 km) | 15.0 | 110 km,<br>90 km,<br>80 km | 98% | 2.1 | ≈15.72% |
| Strategy_5 | ≈0.640606 | 50 | 200.0 km | (25 km,<br>225 km) | 15.7 | 110 km,<br>90 km,<br>80 km | 98% | 0.1 | ≈2.6242038% |
| Strategy_5 | ≈0.829442 | 50 | 345.0 km | (23 km,<br>184 km) | 15.4 | 110 km,<br>90 km,<br>80 km | 98% | 3.5 | ≈24.272727% |
| Strategy_5 | ≈1.16390 | 50 | 960.0 km | (20 km,<br>160 km) | 14.2 | 110 km,<br>90 km,<br>80 km | 98% | 9.4 | ≈66.873239% |
| Strategy_5 | ≈0.25351 | 50 | 270.0 km | (45 km,<br>405 km) | 14.1 | 120 km,<br>110 km,<br>90 km,<br>80 km | 98% | 0.3 | ≈4.0851% |
| Strategy_5 | ≈0.405311 | 50 | 360.0 km | (60 km,<br>540 km) | 13.2 | 120 km,<br>110 km,<br>90 km,<br>80 km | 98% | 5.3 | ≈41.34848% |
| Strategy_5 | ≈0.652449 | 50 | 375.0 km | (25 km,<br>200 km) | 16.2 | 120 km,<br>110 km,<br>90 km,<br>80 km | 98% | 1.5 | ≈11.074074% |
| Strategy_5 | ≈0.934170 | 50 | 697.0 km | (17 km,<br>119 km) | 14.6 | 120 km,<br>110 km,<br>90 km,<br>80 km | 98% | 3.0 | ≈22.136986% |
| Strategy_5 | ≈1.11291 | 50 | 960.0 km | (15 km,<br>135 km) | 16.1 | 120 km,<br>110 km,<br>90 km,<br>80 km | 98% | 7.0 | ≈44.608696% |
| Strategy_5 | ≈0.344096 | 50 | 480.0 km | (80.0 km,<br>640.0 km) | 16.2 | 200 km,<br>120 km,<br>110 km,<br>90 km,<br>80 km | 98% | 2.3 | ≈15.91358% |
| Strategy_5 | ≈0.431319 | 50 | 375.0 km | (75.0 km,<br>675.0 km) | 14.7 | 200 km,<br>120 km,<br>110 km,<br>90 km,<br>80 km | 98% | 2.6 | ≈19.33333% |
| **Strategy_5** | **≈0.652398** | **50** | **550.0 km** | **(50.0 km, 400.0 km)** | **15.3** | **200 km,<br>120 km,<br>110 km,<br>90 km,<br>80 km** | **98%** | **2.6** | **≈18.65359%** |
| **Strategy_5** | **≈0.880917** | **50** | **600.0 km** | **(40.0 km, 360.0 km)** | **14.7** | **200 km,<br>120 km,<br>110 km,<br>90 km,<br>80 km** | **98%** | **1.8** | **14.0%** |
| **Strategy_5** | **≈1.12557** | **50** | **1425.0 km** | **(25.0 km, 250.0 km)** | **14.0** | **200 km,<br>120 km,<br>110 km,<br>90 km,<br>80 km** | **98%** | **4.1** | **≈30.7%** |

**Table 5: Aggregation of optimal solution for each generated network**

Ordinary Least Square Linear Regression analysis (Figure 26) of Table 5 where dependent variable is generated weighted edge location network's small worldness and independent variables are generated weighted edge location network diameter and sum of IADS missile batteries interceptor missile ranges. Here, R-squared is ≈0.7, coefficient of generated weighted edge location network diameter is 0.0008, coefficient of sum of IADS missile batteries interceptor missile ranges is -0.0007 and constant is 0.5615

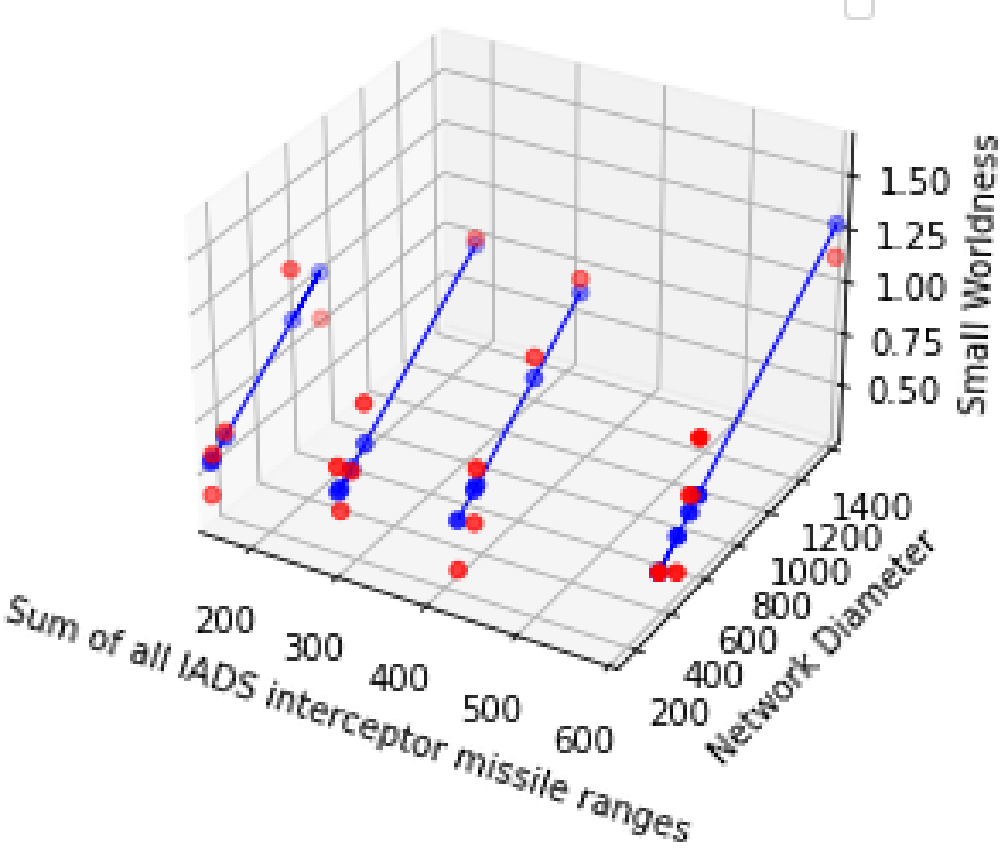

**Figure 26: 3D Depiction of Table 5's linear regression analysis for each sum of IADS missile batteries interceptor missile ranges where sum of IADS missile batteries interceptor missile ranges are sequentially 150 km, 280 km, 400 km and 600 km. Red dots are training data, blue dots are predicted data and blue lines are linear regression lines for each sum of IADS missile batteries interceptor missile ranges.**

Ordinary Least Square Linear Regression analysis (Figure 27) of Table 5 where dependent variable is total unprotected asset value percentage and independent variables are generated weighted edge location network diameter and sum of IADS missile batteries interceptor missile ranges. Here, R-squared is ≈0.6, coefficient of generated weighted edge location network diameter is 0.0463, coefficient of sum of IADS missile batteries interceptor missile ranges is -0.0440 and constant is 15.6197

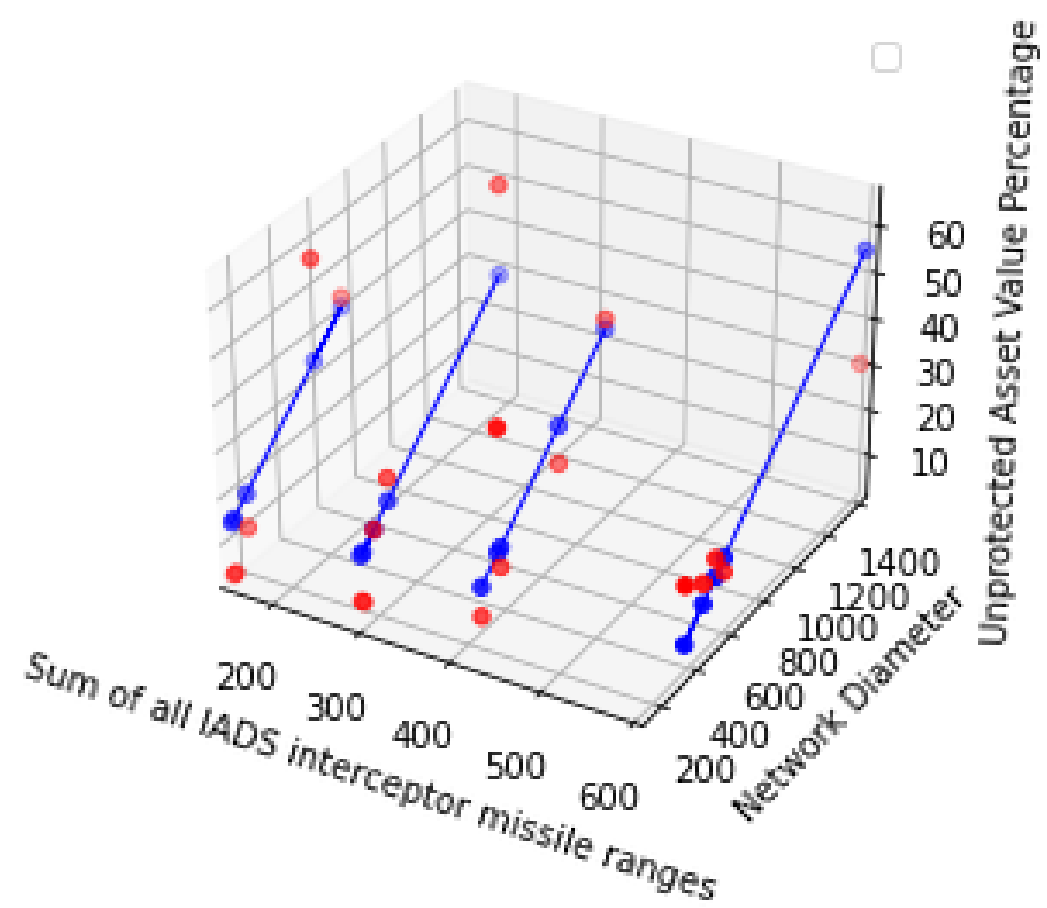

**Figure 27: 3D Depiction of Table 5's linear regression analysis for each sum of IADS missile batteries interceptor missile ranges where sum of IADS missile batteries interceptor missile ranges are sequentially 150 km, 280 km, 400 km and 600 km. Red dots are training data, blue dots are predicted data and blue lines are linear regression lines for each sum of IADS missile batteries interceptor missile ranges.**

## VI. Conclusion

As it is found from computational results that Strategy_5 is the most optimal strategy for small world networks and networks which are not small world networks. In some cases, Strategy_4, Strategy_6 and Strategy_7 are also optimal strategies along with Strategy_5. Eventually, by analyzing computational results, it can be stated that Strategy_6 and Strategy_7 tend to be optimal when the location network has less small world characteristics and the diameter is far longer than the sum of all IADS missile batteries interceptor missile ranges. On the other hand, Strategy_4 tends to be optimal when the location network has more small world characteristics and the diameter is equal or less than the sum of all IADS missile batteries interceptor missile ranges. From regression analysis, it can be stated that, with increasing the location network diameter and the sum of all IADS missile batteries interceptor missile ranges, small worldness value of the location network for optimal strategy also increases (more small world value means less small world characteristics) and, with increasing the location network diameter and the sum of all IADS missile batteries interceptor missile ranges, total unprotected asset value percentages (worst case scenario) of the location network for optimal strategy also increases.

For future research, we may consider scale-free networks [26] as location networks and in place of regression analysis we may use other supervised learning methods such as artificial neural networks [27] to understand relationship patterns among location networks properties and IADS missile battery's interceptor missile's properties.